\documentstyle[12pt,epsfig]{article}

\newcommand{\be}{\begin{equation}}
\newcommand{\ee}{\end{equation}}
\newcommand{\bea}{\begin{eqnarray}}
\newcommand{\eea}{\end{eqnarray}}

\newcommand{\beq}{\begin{equation}}
\newcommand{\eeq}{\end{equation}}

\newcommand{\nn}{\nonumber}

\def\fun#1#2{\lower3.6pt\vbox{\baselineskip0pt\lineskip.9pt
\ialign{$\mathsurround=0pt#1\hfil##\hfil$\crcr#2\crcr\sim\crcr}}}

\begin{document}

\title{Two-pion decays of mesons and confinement
     singularities}

\author{A.V. Anisovich,   V.V. Anisovich, M.A. Matveev, V.A. Nikonov,
          J. Nyiri,\\ and A.V. Sarantsev}

\date{\today}

\maketitle

\begin{abstract}
We consider the two-pion decay of the $\rho$-meson, the $^3S_1q\bar
q$-state of the constituent quarks -- the decay being determined by
the transition $q\bar q\to \pi\pi$ contains information about
confinement interactions. One can specify in this decay two types of
transitions: ({\bf i}) the bremsstrahlung radiation of a pion $q\to
q+\pi$ (or $\bar q\to\bar  q+\pi$) with the subsequent fusion
$q\bar q\to \pi$, and ({\bf ii}) the direct transition $q\bar q\to
\pi\pi$. We demonstrate how in the amplitudes of the corresponding
transitions the quark singularities have to disappear, {\it i.e.}
what is the way the quark confinement at relatively short distances
can be realized. We calculate and estimate the contributions of
processes with bremsstrahlung radiation of the pion and of the
direct transition $q\bar q\to \pi\pi$. The estimates demonstrate
that the processes involving the direct transition $q\bar q\to
\pi\pi$ are necessary, but they cannot be determined unambiguously
by the decay $\rho(775)\to \pi\pi$. We conclude that for the
determination of the $q\bar q\to \pi\pi$ transition more complete
data on the resonance decays into the $\pi\pi$ channels are required
than those available at the moment.

\end{abstract}

\section{Introduction}

We make an attempt to restore the structure of hadrons (beginning,
naturally, with the simplest ones -- the light mesons) using notions
like constituent quarks and effective (massive) gluons. So, we are
working with effective particles and build effective interactions
and effective Hamiltonians which can give an adequate description of
the region of soft interactions of quarks and gluons (see
\cite{book3}, Chapters 9 and 10).

Recently we succeeded in constructing spectral integral equations
for single-component quark-antiquark systems ($b\bar b$, $c\bar c$
\cite{SI-0,SI-bb,SI-cc} and light quark $q\bar q$ states with
isospin $I=1$ or quarkonium type ones like $\phi$ and $\omega$
\cite{SI-qq}). We found interactions describing both the levels of
these states and their radiative decays (see also \cite{book3}).

Considering mesons consisting of light quarks \cite{SI-qq}, we
obtained series of radially excited states. These states form with a
good accuracy linear trajectories with a universal slope on the
($n,M^2$)-planes, where $M^2$ and $n$ are masses squared and radial
quantum numbers, respectively. The systematization of meson states
on ($n,M^2$) planes leads to a good agreement with the data
\cite{syst}. Examples for states lying on linear trajectories are
[$\rho(775)$, $\rho(1460)$, $\rho(1870)$, $\rho(2110)$],
[$\omega(780)$, $\omega(1430)$, $\omega(1830)$, $\omega(2205)$],
[$\pi(140)$, $\pi(1300)$, $\pi(1800)$, $\pi(2070)$, $\pi(2360)$] and
many others (see \cite{book3} for more detail).

In the present paper we investigate hadronic decays of mesons. We
start with the simplest one, $\rho\to\pi\pi$. Considering
simultaneously the decays of the $^3S_1q\bar q$ mesons, namely
$\omega\to\gamma\pi$, $\rho\to\gamma\pi$ and $\rho\to\pi\pi$, we
have the possibility to study the confinement singularities owing to
quark exchange (the Gribov confinement singularity).

In the radiative decay reactions we face two mechanisms:\\
{\bf(i)} a bremsstrahlung emission of a photon $q\to \gamma+q$ with
a subsequent transition $q\bar q\to \pi$ and \\
{\bf(ii)} a bremsstrahlung-type emission of a pion $q\to \pi+q$ with
a subsequent annihilation $q\bar q\to \gamma$.

In the $\rho\to\pi\pi$ reaction,\\
{\bf(i)} along with the emission of a pion $q \to \pi +q$ and the
subsequent transition $q\bar q\to \pi$,\\
{\bf(ii)} we observe a decay caused by Gribov's confinement
mechanism.\\
We see that the  $q\bar q\to \pi\pi $ transition is necessary for
the description of the $\rho(775)\to\pi\pi$ decay width. However,
the structure of the $q\bar q\to \pi\pi$ amplitude itself is not
determined unambiguously, there remains a freedom of choice for the
form of the singularity. We consider here some possible versions.
Doing so, on the basis of experimental data for $\rho(775)\to\pi\pi$
we estimate the contribution of the amplitude $q\bar q\to \pi\pi$ in
different versions of the interaction.

In order to convince ourselves that the decay amplitude
$\rho(775)\to\pi\pi$ does not contain quark singularities in any of
the considered versions of the ($q\bar q\to \pi\pi$)-interaction
({\it i.e.} in all these cases the quarks are confined), we
investigate the two-channel spectral equation for the $\rho$-meson,
taking into account the $q\bar q$ and $\pi\pi$ channels. Supposing
that the non-resonance low-energy $\pi\pi$-interaction is small, we
give a solution for the two-channel spectral integral equation. As
it turns out, in this case the $\rho$-meson propagator obtains a
self-energy part $[M^2_\rho - s]^{-1}\to [M^2_\rho - s- B(s)]^{-1}$,
where $B(s)$ is the amplitude of the $\rho(775)\to q\bar q\to
\pi\pi\to  q\bar q\to\rho(775)$ transition. Our calculations show
explicitly that the self-energy amplitude $B(s)$ does not contain
threshold singularities (and imaginary parts) related to the $q\bar
q$-channels; in other words, it describes confined quarks. It can be
seen also that the quark confinement is due to the strong
singularities in the $q\bar q\to q\bar q$ - singularity channel,
which create the $V_{confinement}(r)\sim b\, r$ barrier
\cite{book3,SI-qq}. However - and this can also be seen explicitly -
the $q\bar q\to \pi\pi$ transition plays an essential role in
confining the quarks. If it were not present, the bremsstrahlung
emission of pions would destroy the $\rho(775)$ meson quickly,
giving $\Gamma_{\rho(775)\to\pi\pi}(bremsstrahlung\, emission\, of\,
pion) \simeq 2000$ MeV.

The structure of the paper is as follows. In the Introduction we
provide a short description of those basic statements which we
assume to be known (see \cite{book3}) and from which we start our
investigations of the $\rho\to\pi\pi$ reaction. In Section 2 we
calculate diagrams with bremsstrahlung emissions of pions. Then, in
Section 3, we discuss the confinement singularities and calculate
the triangle diagram owing to the direct production process $q\bar
q\to \pi$. In Section 4 we discuss two-channel ($q\bar q$ and
$\pi\pi$) equations for $\rho$-meson states as well as the
self-energy part in by the transition $\rho(775)\to q\bar q\to
\pi\pi\to  q\bar q\to\rho(775)$. Results of the calculations are
summarized in the Conclusion.

\subsection{Spectral integral equation and confinement singularity}

\begin{figure}[ht]
  \centerline{\epsfig{file=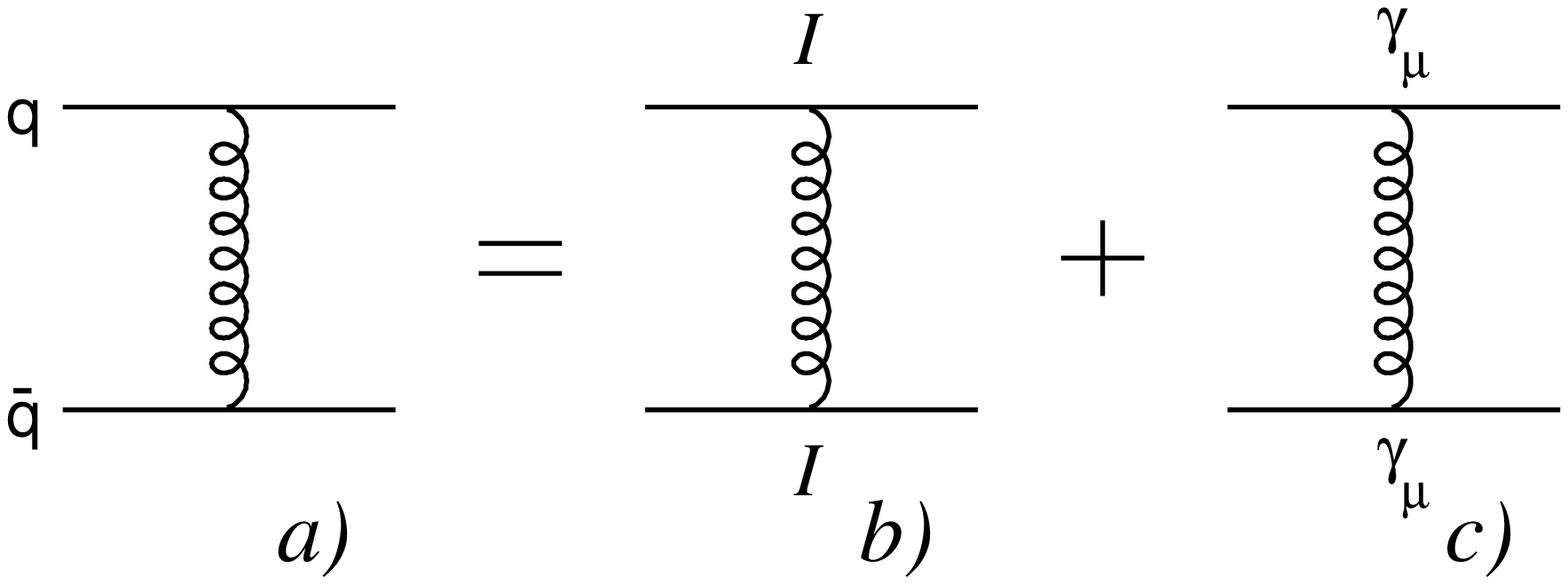,width=9cm}}
  \centerline{\epsfig{file=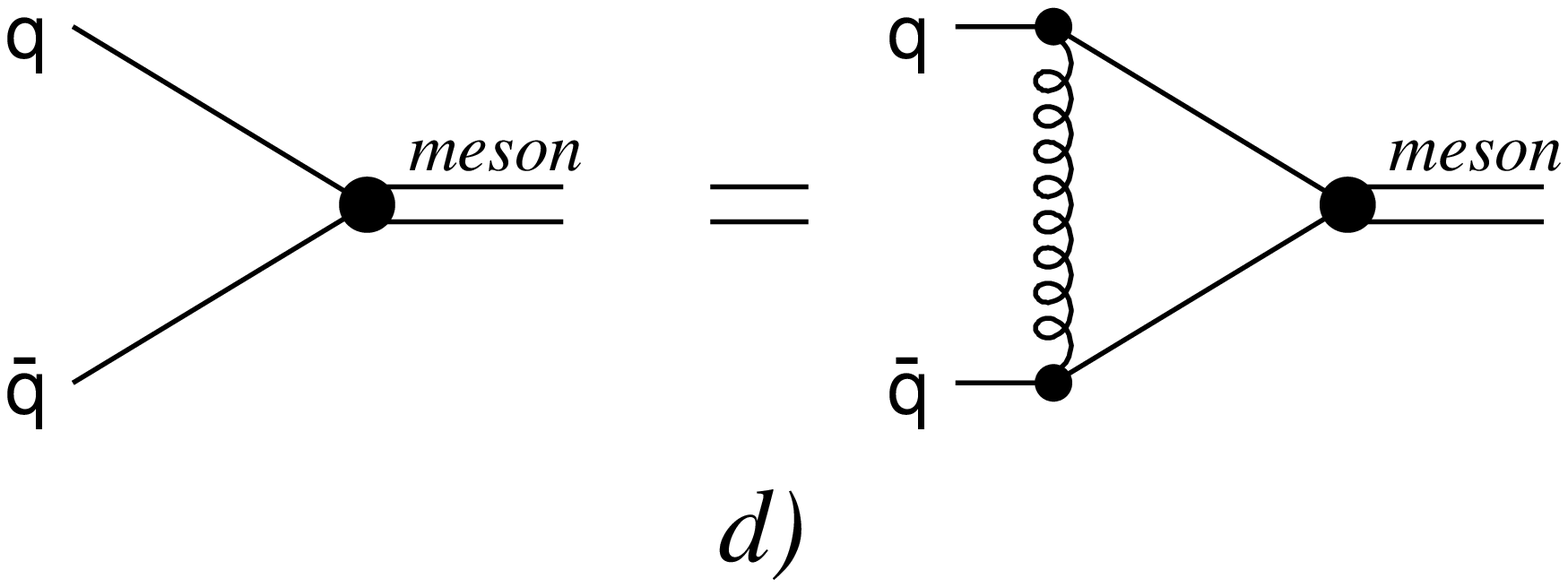,width=9cm}}
      \caption{Diagrams for confinement singularities
  $1/q^4$: scalar (b) and vector (c) exchanges in the $t$-channel.
    (d) Graphical representation of the spectral integral equation for
    the meson-$q\bar q$ vertex with $t$-channel
    confinement interaction.
  \label{pic_1}}
  \end{figure}

The linearity of the trajectories in the ($n,M^2$) planes
(experimentally -- up to large $n$ values, $n\le 7$) provides us the
$t$-channel singularity $V_{conf}\sim 1/q^4$ or, in coordinate
representation, $V_{conf}\sim r$. In the coordinate representation
the confinement interaction can be written in the following
potential form \cite{SI-qq} (see Fig. \ref{pic_1}):
 \bea \label{I1}
  &&V_{conf}=(I\otimes I)\,b_S\,r + (\gamma_\mu\otimes
  \gamma_\mu)\,b_V\,r\ ,
  \\ \nn
  &&b_S\simeq -b_V \simeq 0.15\,\, {\rm GeV}^{-2}\ .
  \eea

  \begin{figure}[h]
  \centerline{\epsfig{file=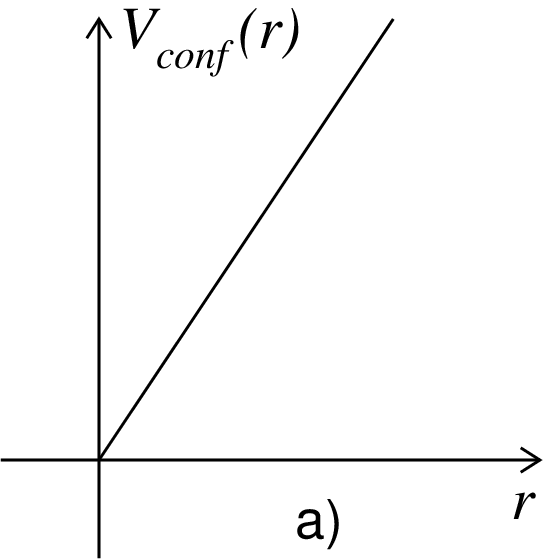,width=4cm} \hspace{2cm}
                 \epsfig{file=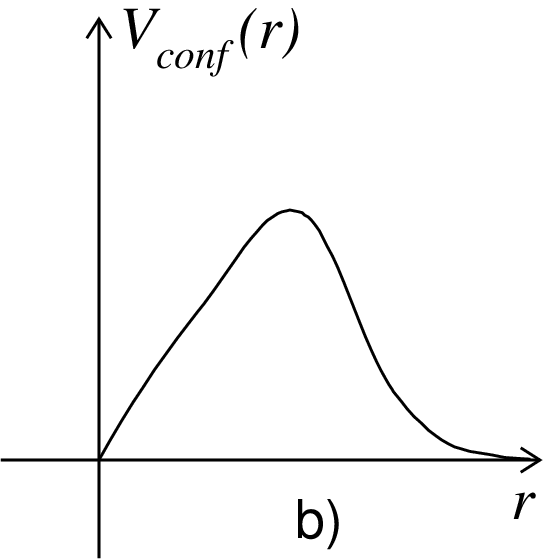,width=4cm}}
      \caption{Confinement singularity in coordinate representation
    a) without cut-off: $V_{conf}(r)\sim b\, r$, and b)
  introducing a cut-off: $V_{conf}(r)\sim b\, r e^{-\mu r}$.
  \label{pic_2}}
  \end{figure}

  \begin{figure}[h]
  \centerline{\epsfig{file=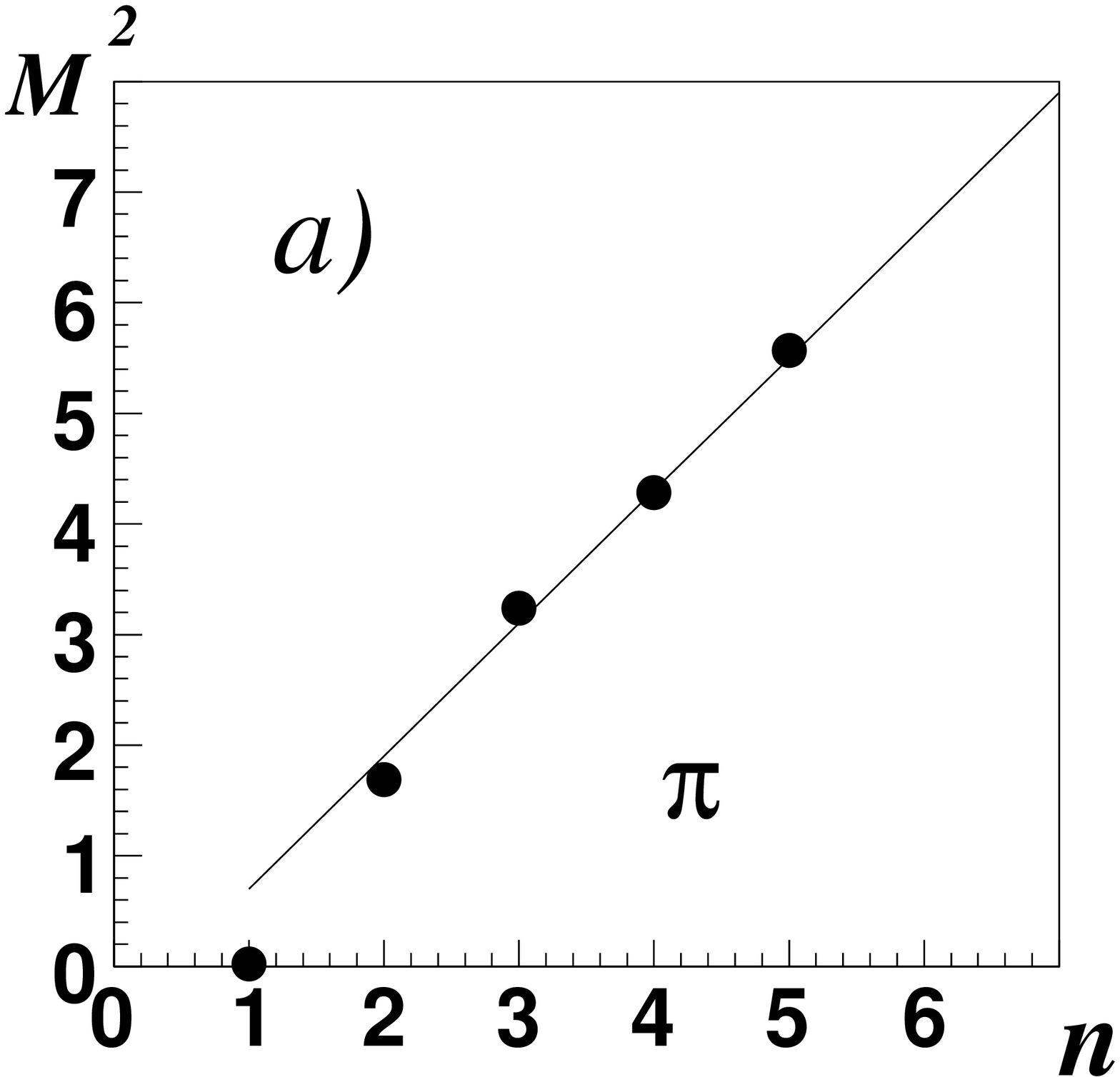,width=4cm}
                 \epsfig{file=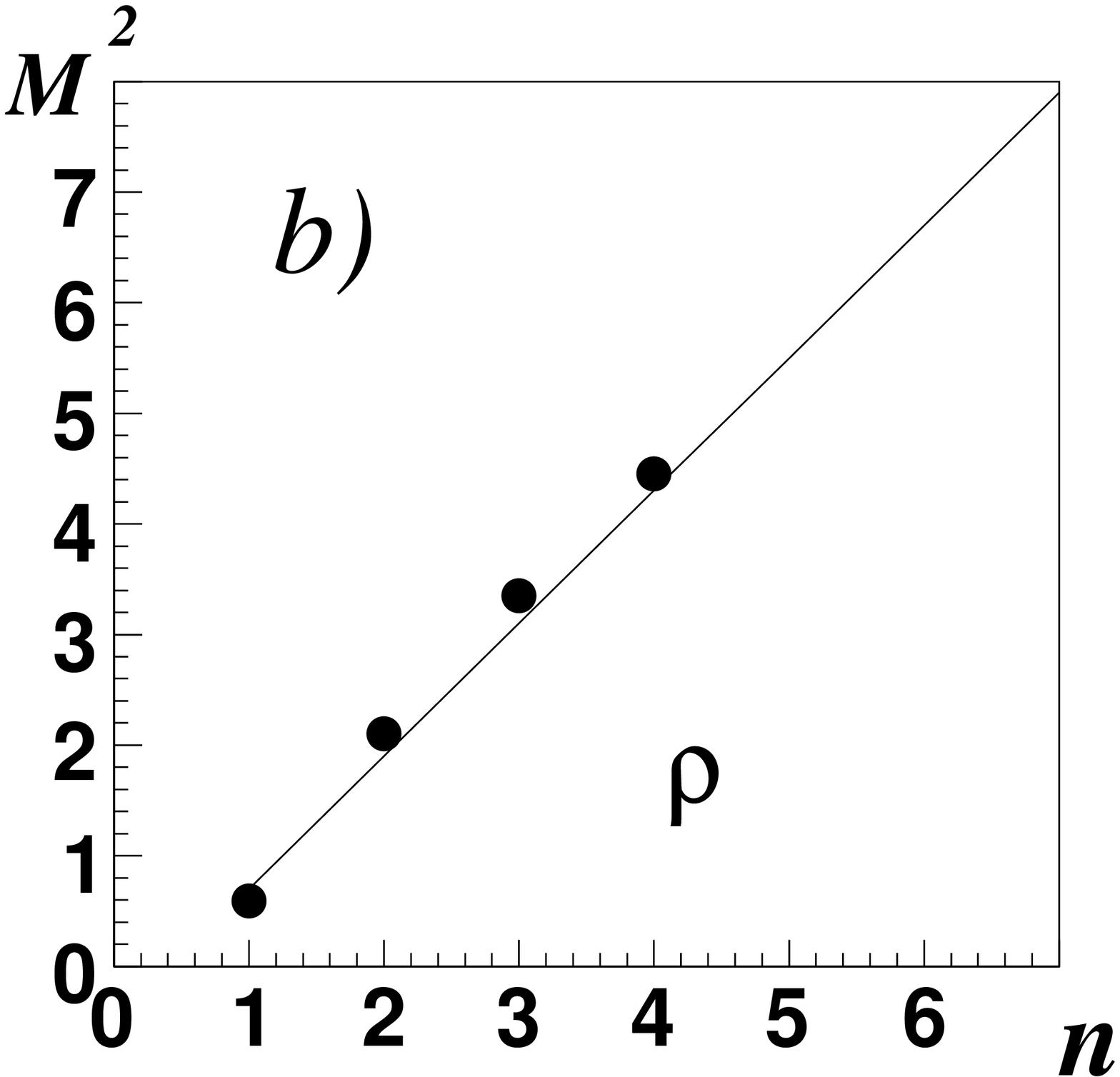,width=4cm}}
  \centerline{\epsfig{file=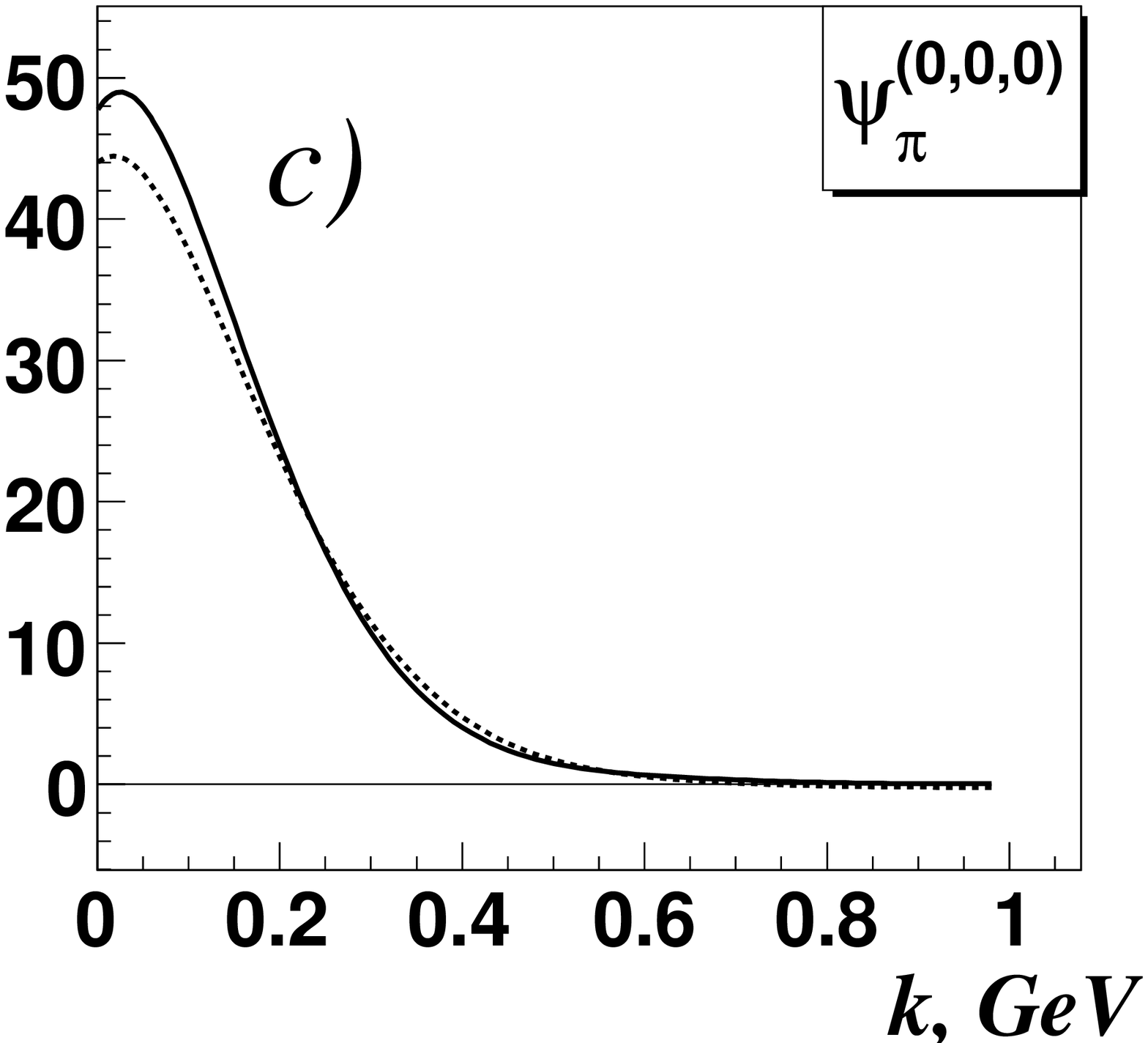,width=4cm}
                 \epsfig{file=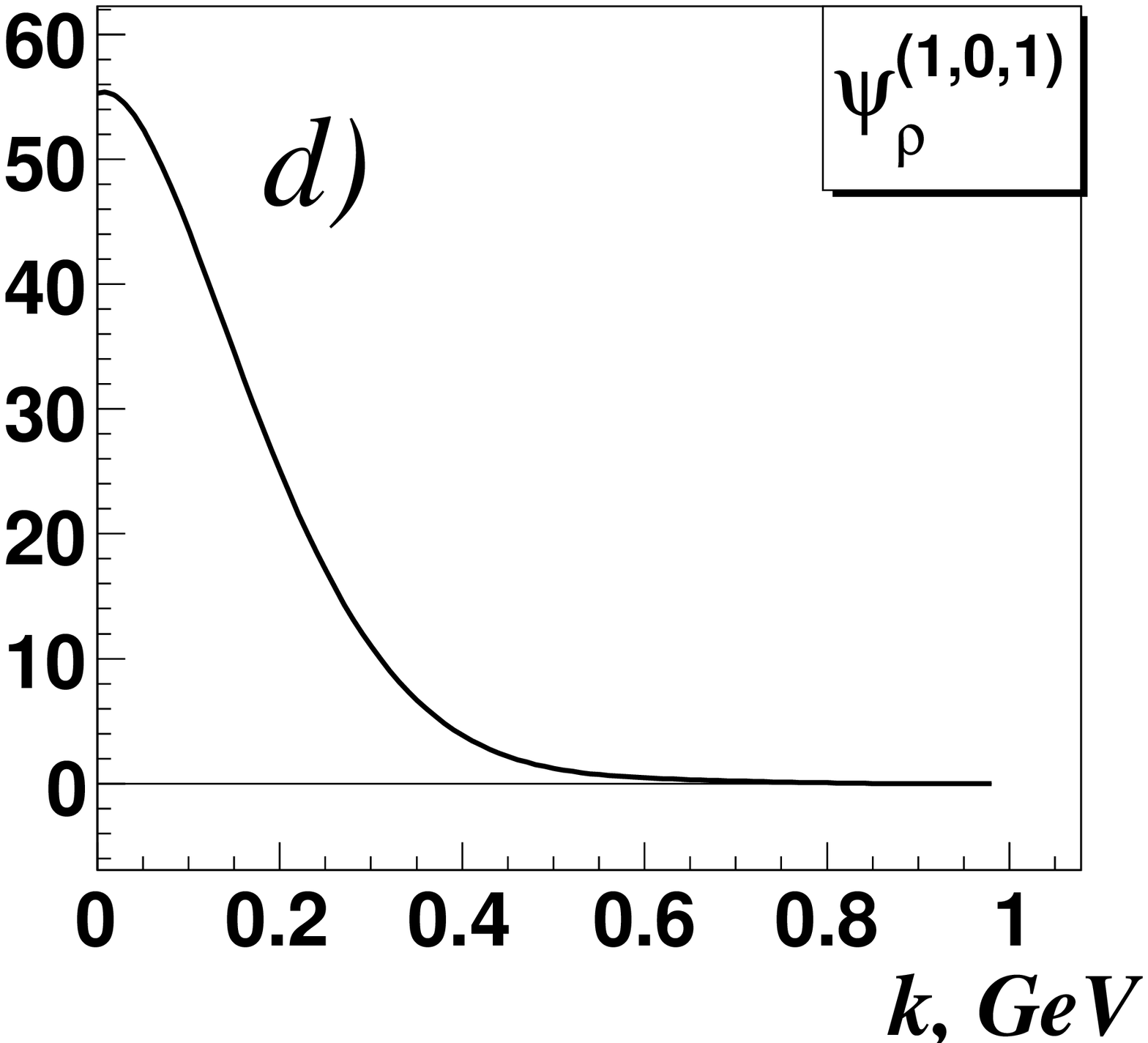,width=4cm}}
      \caption{a) The $\pi$-trajectory in the $(n,M^2)$ - plane;
  b) $\rho /\omega$ - trajectory; c) wave function of the $\pi$-meson
  obtained in the spectral integral equation \cite{SI-qq} (solid line)
  and in the phenomenological fit \cite{dn} (dotted line).
  \label{pic_3}}
  \end{figure}

The position of $q\bar q$ levels and data on radial decays tell us
that singular $t$-channel exchanges are necessary both in the scalar
($I\otimes I$) and the vector ($\gamma_\mu\otimes \gamma_\mu$)
channels. The $t$-channel exchange interactions (\ref{I1}) can take
place both in white and colour states, ${\bf c}={\bf 1}+{\bf 8}$
though, of course, the colour-octet interaction plays a dominant
role.

The spectral integral equation for the meson-$q\bar q$ vertex (or
for the $q\bar q$ wave function of the meson, see Fig. \ref{pic_1}d)
was solved by introducing a cut-off into the interaction (\ref{I1}):
$r\to re^{-\mu r}$, see Fig. \ref{pic_2}. The cut-off parameter is
small: $\mu\sim 1-10$ MeV; if $\mu$ is changing in this interval,
the $q\bar q$-levels  with $n\le 7$ remain practically the same.

In \cite{SI-bb,SI-cc,SI-qq} the spectral integral equations were
solved in momentum representation -- this is natural, since we used
dispersion integration technics (see the discussion in
\cite{book3}). In this representation the interaction is re-written
as
  \bea
  re^{-\mu r}\to
  8\pi\left(\frac{4\mu^2}{(\mu^2-t_{\perp})^3}-\frac{1}{(\mu^2-t_{\perp})^2}
  \right).
  \label{1-1-2}
  \eea
In general, having in mind that in the framework of spectral
integration (as in dispersion technics) the total energy is not
conserved, we have to write
  \bea
  t_{\perp}=(k^\perp_1-k^{'\perp}_1)_\mu(-k^\perp_2+k^{'\perp}_2)_\mu\
  \label{1-1-3}
  \eea
for the momentum transferred, where $k_1$ and $k_2$ are the momenta
of the initial quark and antiquark, while $k'_1$ and $k'_2$ are
those after the interaction. The index $\perp$ means that we use
components perpendicular to the total momentum $p$ for the initial
state and to $p'$ for the final state:
  \bea
  &&
  k^\perp_{i\,\mu}=g^{\perp p}_{\mu\nu}k_{i\,\nu}\ , \quad g^{\perp
  p}_{\mu\nu}=g_{\mu\nu}-\frac{p_\mu p_\nu}{p^2}\ , \quad p=k_1+k_2 \
  , \quad p^2=s\, ,
  \\ \nn
  && k'^{\perp}_{i\,\mu}=g^{\perp p'}_{\mu\nu}k'_{i\,\nu}\ , \quad
  g^{\perp p'}_{\mu\nu}=g_{\mu\nu}-\frac{p'_\mu p'_\nu}{p^{'2}}\ ,
  \quad p'=k'_1+k'_2\, , \quad p'^2=s' \ .
  \label{1-1-4}
  \eea
For quarks with equal masses we write
$t_{\perp}=(k^\perp_1-k^{'\perp}_1)^2=(-k^\perp_2+k^{'\perp}_2)^2$.
If energies in the initial and final states are equal ($s=s'$), we
have $t^\perp=-\vec q\,^2$, and the $t$-channel singular term in the
right-hand side of (\ref{1-1-2}) is equivalent to the interaction
$r\,e^{-\mu r}$ after the Fourier transform. In the general case we
can write
  \bea
  \label{1-1-5}
  r^N e^{-\mu
  r}&=&\int\frac{d^3q}{(2\pi)^3}e^{-i\vec q\vec r} I_N(t_\perp)\ ,
  \\ \nn
  I_N(t_\perp)&=&\frac{4\pi(N+1)!}{(\mu^2-t_\perp)^{N+2}}
  \sum^{N+1}_{n=0}\left(\mu+\sqrt{t_\perp}\right)^{N+1-n}
  \left(\mu-\sqrt{t_\perp}\right)^{n}\ .
  \eea
In Fig. \ref{pic_3}a,b the $(n,M^2)$-trajectories for $\pi$, $\rho$
and $\omega$ mesons are shown (the $\rho$ and $\omega$ trajectories
coincide with a good accuracy). As we see, the $\pi$ meson mass
($\sim 140\, {\rm MeV}$) sticks out on the linear trajectory. This
happens due to an instanton-induced interaction. In Fig.
\ref{pic_3}c,d one can see the pion and $\rho/\omega$ wave
functions. At $k> 0.15$ keV the pion wave function obtained in the
spectral integral equation coincides with a good accuracy with the
phenomenological wave function \cite{dn}, which was found by the fit
of the pion form factor; small deviations ($\sim 10\%$) are observed
in the region of very small $k$ values.

\subsection{Instability of $q\bar q$ levels - quark deconfinement}

All quark-antiquark levels except for the lowest $J^P=0^-$ states
decay into hadron states, {\it i.e.} quarks become deconfined. New
quark pairs are produced and, as a result, the initial quarks leave
the confinement trap, finding a suitable partner among the newly
born quarks to form white hadrons in the final state.

Such general considerations about quark deconfinement were expressed
many times and discussed long ago (see, for example,
\cite{Satz,A74,DTU} and references therein). However, it is the
details of the mechanism what really matters, and they were not
presented. A detailed scheme for the quark deconfinement was
suggested in \cite{Gribov} -- in the main points the present
investigations of the deconfinement process follow these ideas.

\subsubsection{Instability of $q\bar q$ levels owing to radiative
  decays.}

In the calculations \cite{SI-qq} the levels are instable, as can be
seen from Fig. \ref{pic_2}b. But this is, so to say, a ``technical''
instability, owing to the method of calculation. It is convenient to
discuss the physical mechanism of the instability, considering first
the radiative decays.

  \begin{figure}[h]
  \centerline{\epsfig{file=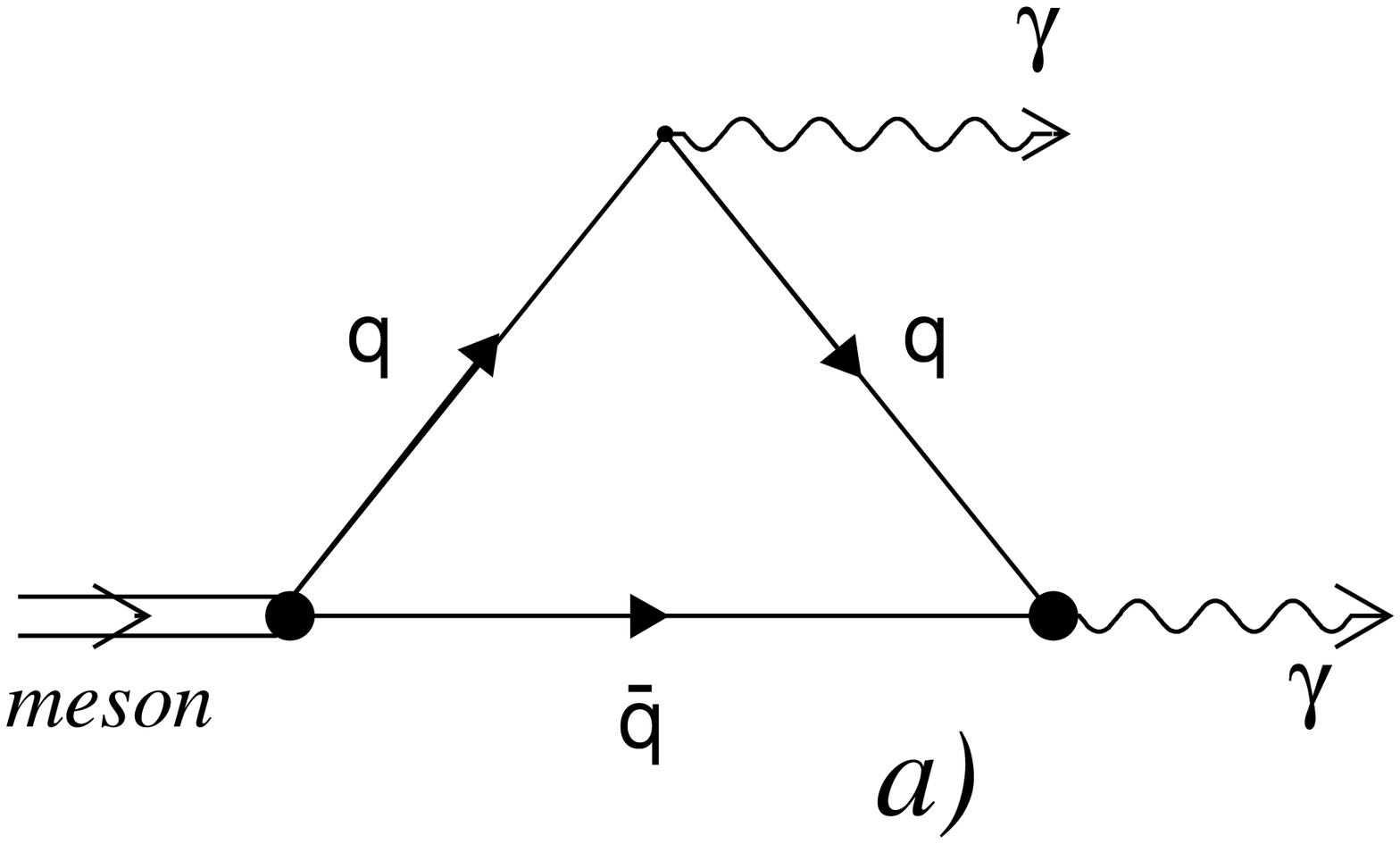,width=5cm}
                 \epsfig{file=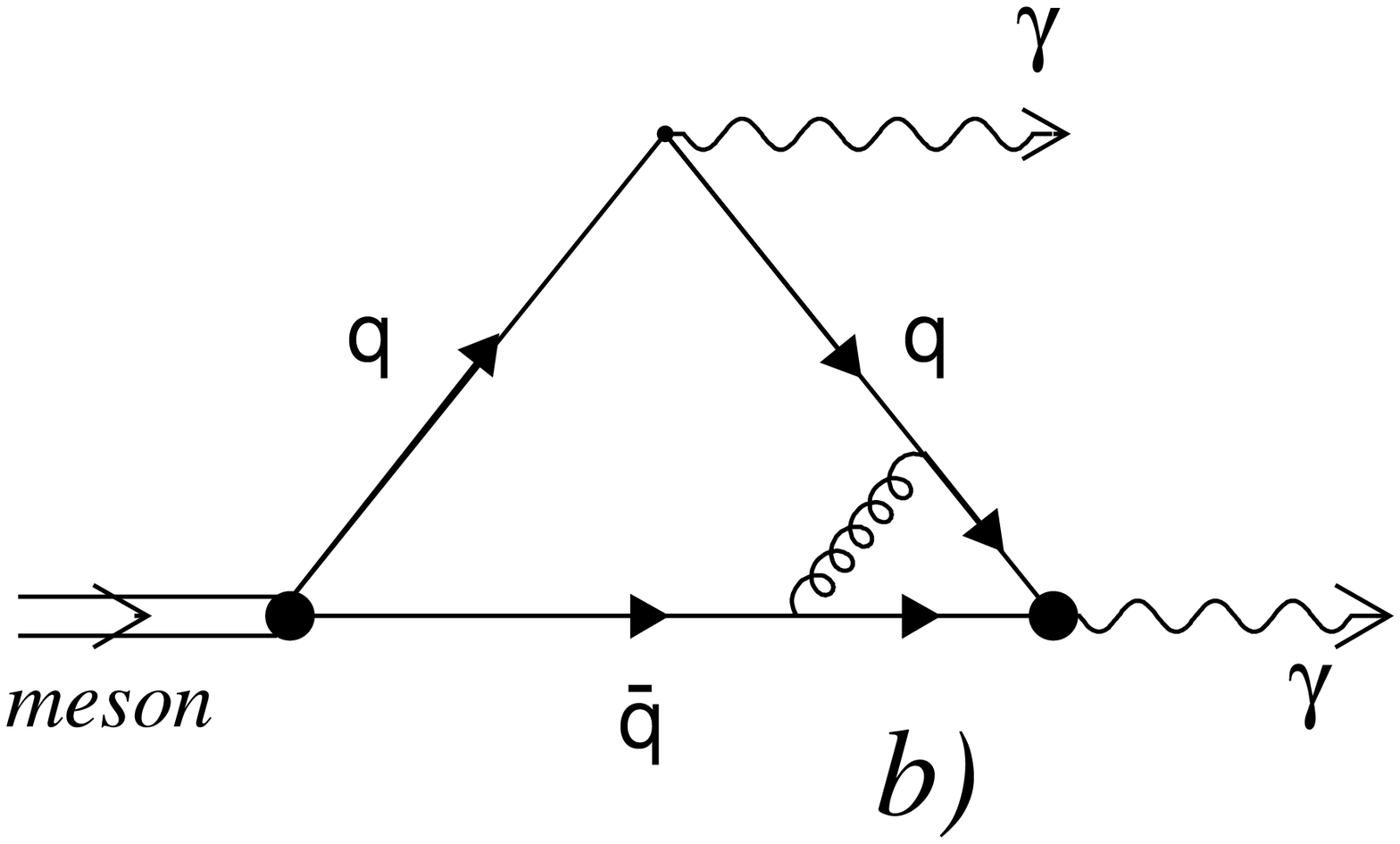,width=5cm}}
  \caption{Quark model mechanism for the $meson\to\gamma\gamma$ decay:
  a) bremsstrahlung photon radiation and the subsequent annihilation
  process $q\bar q\to\gamma$.  The annihilation block includes quark
  interactions, among others long range ones (b). \label{pic_4}}
  \end{figure}

{\bf (i) $meson\to\gamma\gamma$ decays}\\
Considering this decay in the framework of the quark model, we face
two relevant mechanisms: the bremsstrahlung radiation of a photon by
a quark or antiquark, and the subsequent annihilation $q\bar
q\to\gamma$. In Fig. \ref{pic_4} the bremsstrahlung radiation by a
quark is shown only, but, naturally, there is a similar process of
bremsstrahlung radiation by an antiquark. In the annihilation
vertex, quark interactions (including the confinement interaction)
are also present. In the considered process the quarks do not leave
the quark trap, but, on the other hand, mesons are also not flying
away. So, let us consider a different decay process in which a meson
(a pion) leaves the confinement region.

  \begin{figure}[h]
  \centerline{\epsfig{file=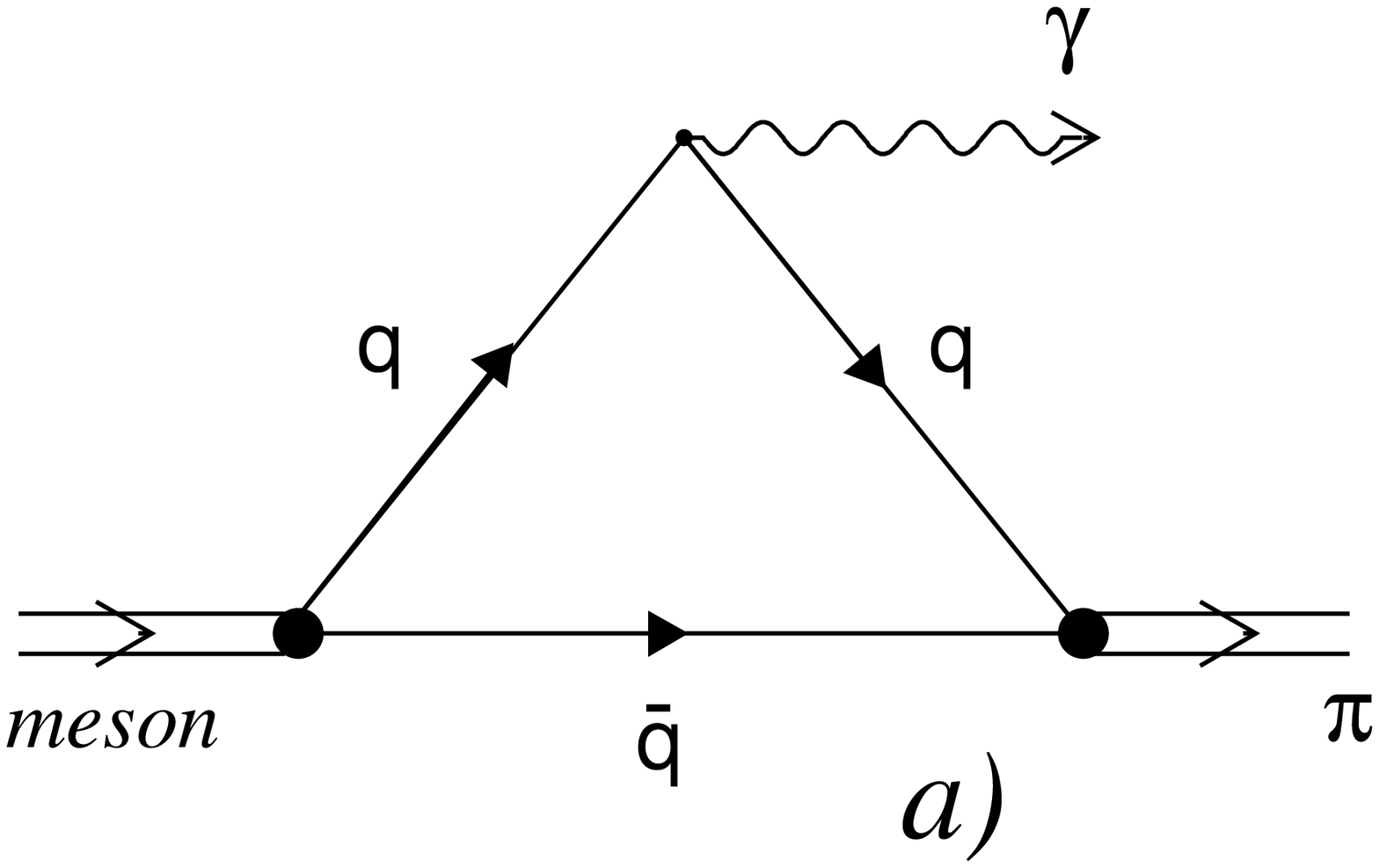,width=5cm}
                 \epsfig{file=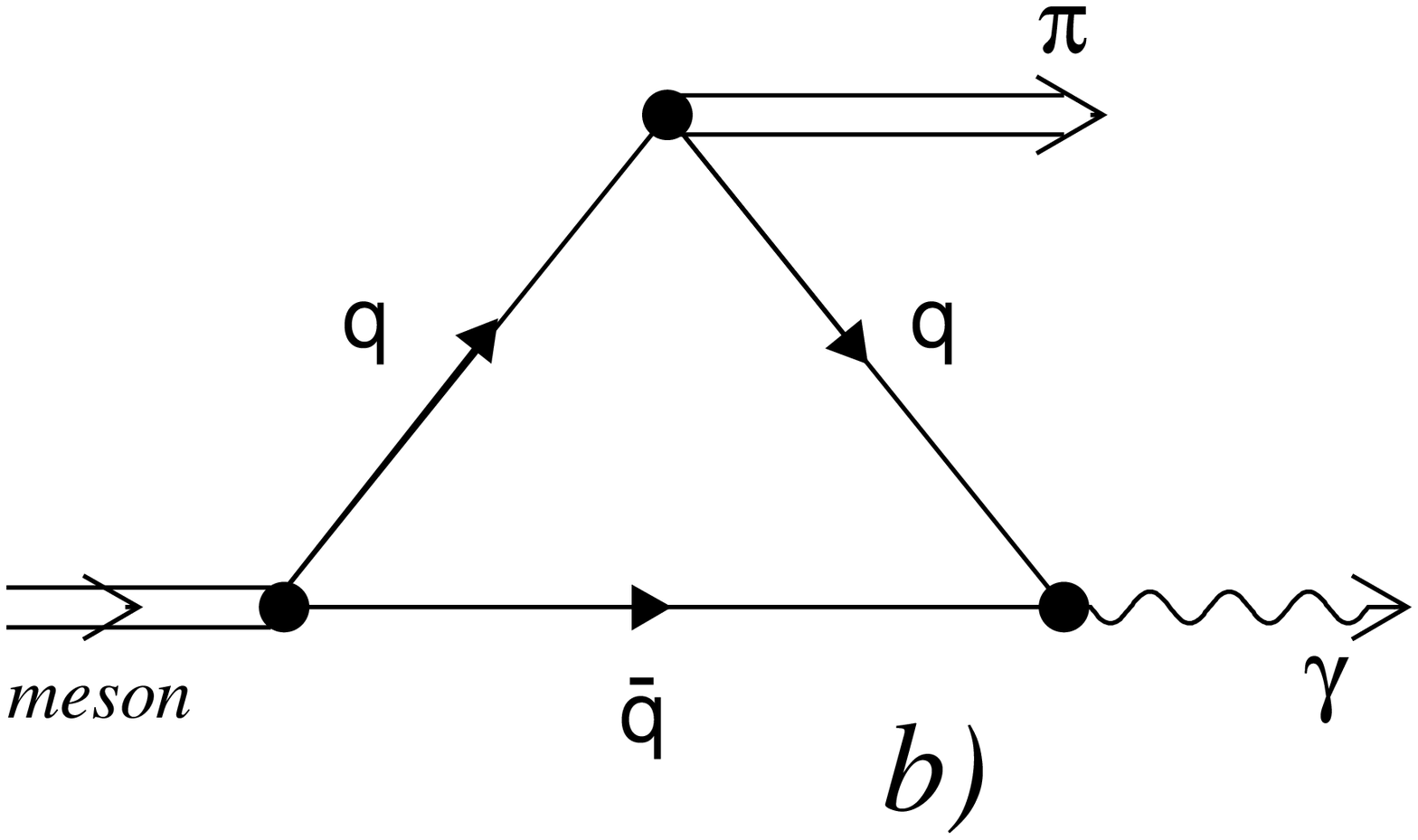,width=5cm}}
  \centerline{\epsfig{file=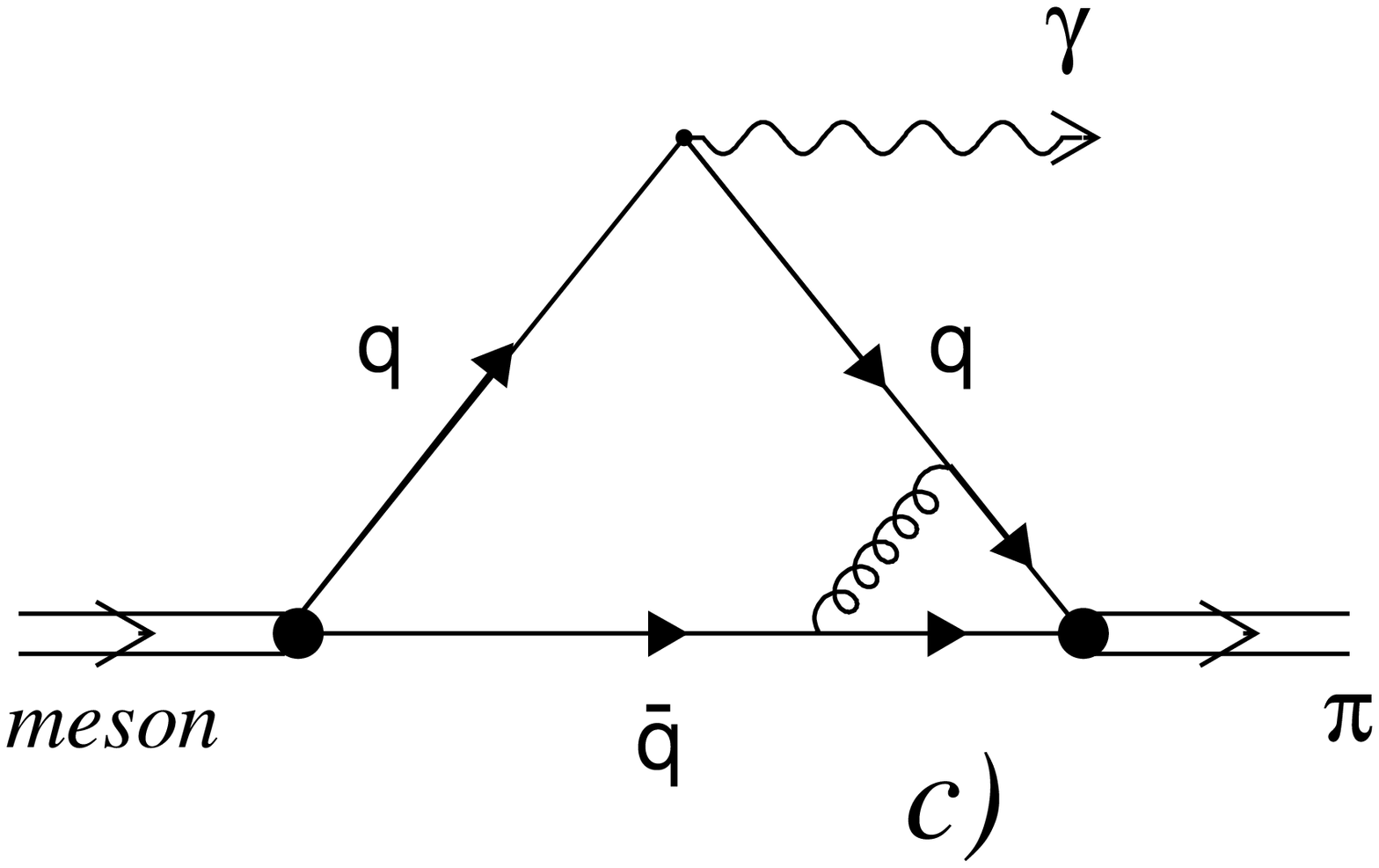,width=5cm}
                 \epsfig{file=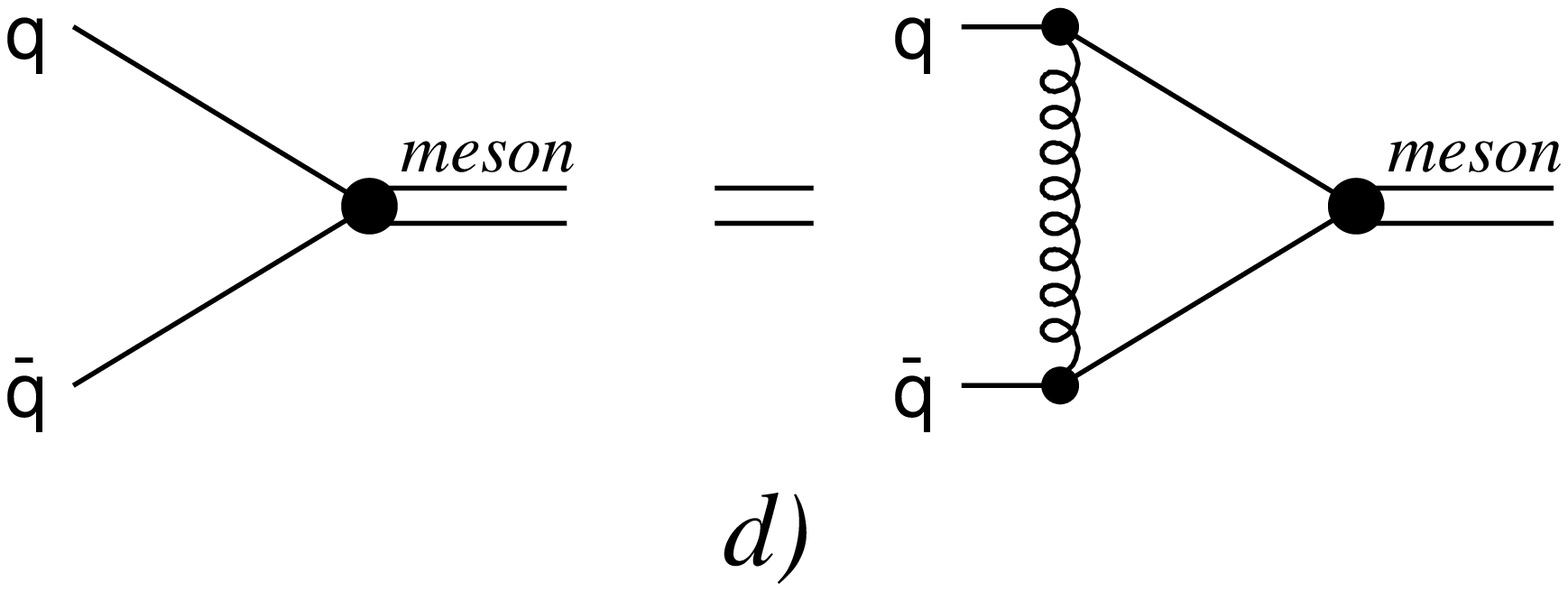,width=6cm}}
  \caption{Emission of a photon (a) and a pion (b) in the
  $\rho,\omega\to\gamma\pi$ processes. c) The annihilation block
  $q\bar q\to\pi$ includes quark interactions similar to that in Fig.
  4b.
    d) Spectral integral
  equation which determines the $q\bar q\to\pi$ transition block.
  \label{pic_5}}
  \end{figure}

{\bf (ii) $meson\to \gamma\pi$ decays}\\
This decay consists of two types of processes: the bremsstrahlung of
a photon with the subsequent $q\bar q\to \pi$ transition (Fig.
\ref{pic_5}a) and the bremsstrahlung-type pion radiation (the pion
has a small mass) with the subsequent $q\bar q\to \gamma$
annihilation (Fig. \ref{pic_5}b) (as in the previous case, we show
here the radiation by a quark only). In Fig. \ref{pic_5}c we
emphasize that in the annihilation block $q\bar q\to \pi$ the
quark-antiquark interactions are included. It is just due to these
interactions that the $q\bar q\to \pi$ block satisfies the spectral
integral equation \cite{SI-qq} (see Fig. \ref{pic_5}d); the spectral
integral equation contains, naturally, the confinement interaction.

Hence, in both processes, $meson\to \gamma\gamma$ (Fig. \ref{pic_4})
and $meson\to\gamma\pi$ (Fig. \ref{pic_5}), the confinement
interaction (or the confinement singularity) plays a relevant role
at the last stage: in the $q\bar q$ annihilation and in the
formation of a pion.

{\bf (iii) The $\rho\to\pi\pi$ decay}\\
In hadronic decays the confinement interaction plays, of course, a
decisive role but the brems\-strah\-lung-type emission of pions may
also be important if there are pions among the particles of the
outgoing state.

  \begin{figure}[h]
  \centerline{\epsfig{file=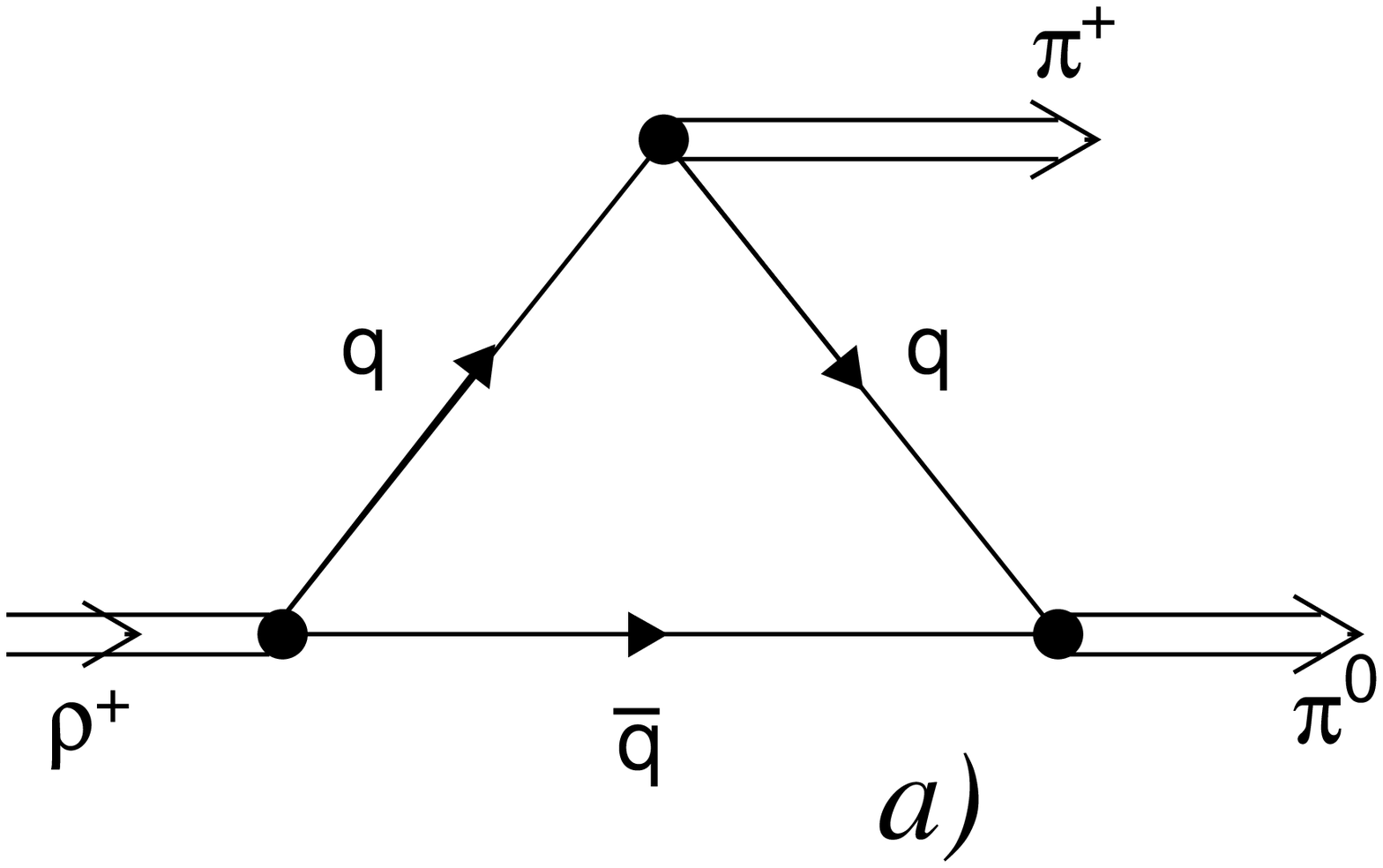,width=4cm}
                 \epsfig{file=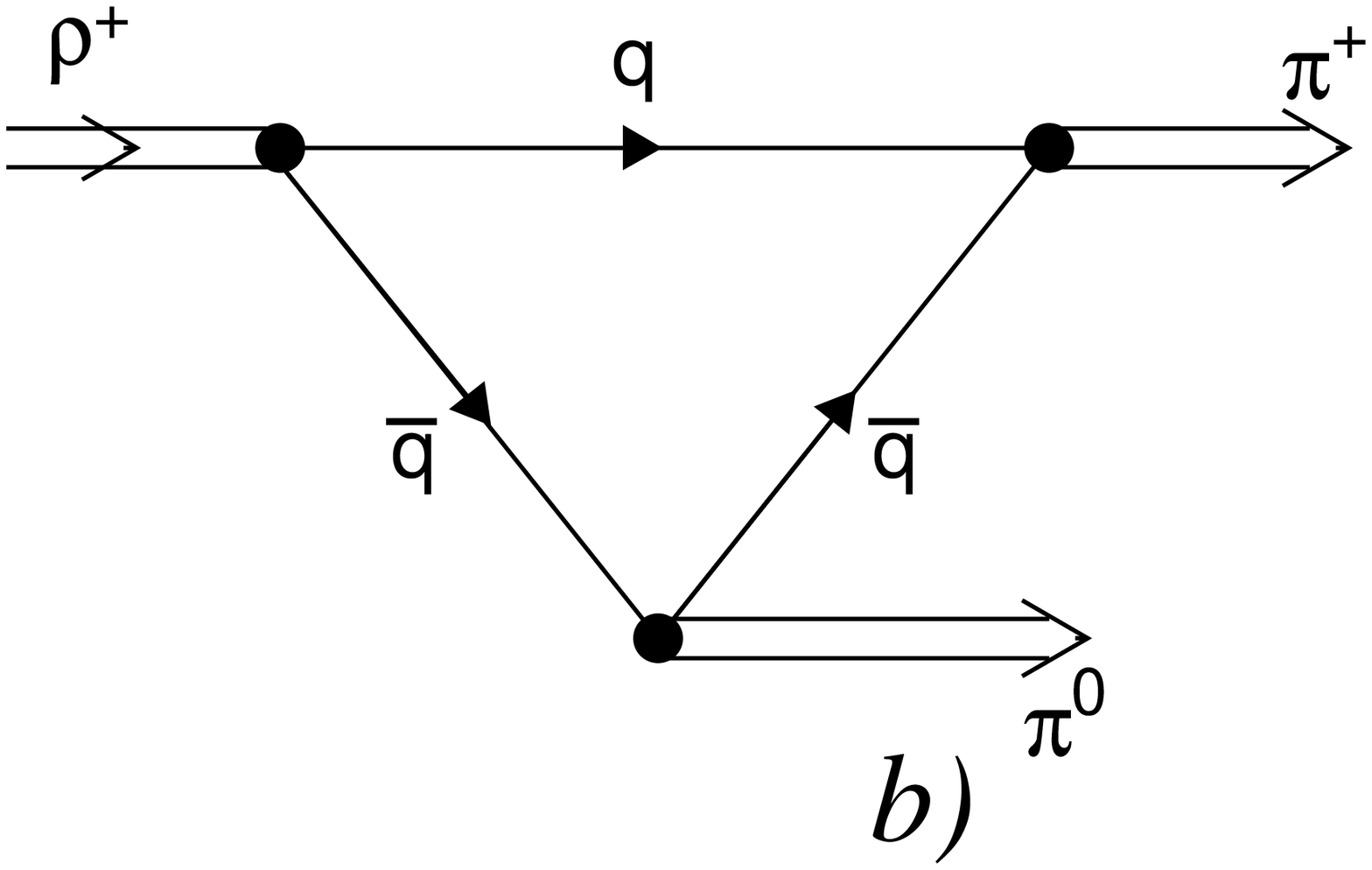,width=4cm}}
  \centerline{\epsfig{file=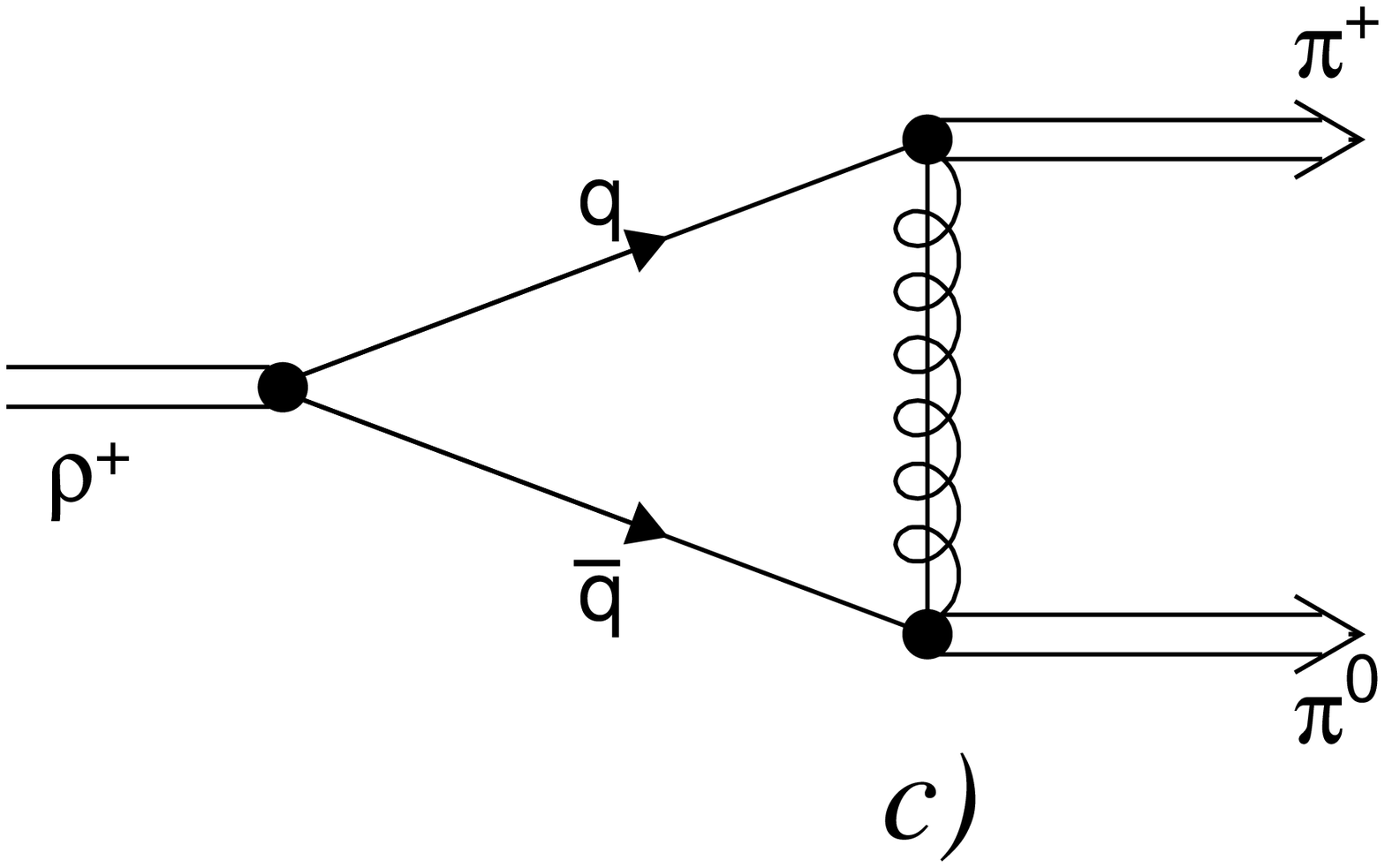,width=4cm}}
      \caption{The $\rho^+\to\pi^+\pi^0$ process: a,b) with a
  bremsstrahlung-type pion emission and c) with the $q\bar q\to\pi\pi$
  transition, realized owing to the $t$-channel confinement
  singularity with fermion quantum numbers.
  \label{pic_6}}
  \end{figure}

The decay processes $\rho\to\pi\pi$ (to be definite, we consider
$\rho^+\to\pi^0\pi^+$) are demonstrated in Fig. \ref{pic_6}. While
the processes Fig. \ref{pic_6}a,b do not differ essentially from
those shown in Fig. \ref{pic_5}a,b for $\rho\to\gamma\pi$, the
process Fig. \ref{pic_6}c is principally different. In the processes
in Fig. \ref{pic_5}a,b and Fig. \ref{pic_6}a,b one of the particles
(the photon or the pion) is emitted by a constituent quark when it
is inside a ``bag'' (in the region where colour quarks can exist).
Further, two quarks form a particle: a pion (Fig. \ref{pic_5}a and
\ref{pic_6}a,b) or a photon (Fig. \ref{pic_5}b) -- obviously, with
the participation of the confinement interaction (see Figs.
\ref{pic_4}b and \ref{pic_5}c). In the process shown in Fig.
\ref{pic_6}c there is a transition of $q\bar q$ into two mesons,
$q\bar q\to \pi\pi$. The confinement singularity, but with fermion
quantum numbers, has to play a crucial role here. As it was stressed
by Gribov \cite{Gribov}, the corresponding forces must be long-range
ones, {\it i.e.} the effective mass of the exchange fermion state
has to be small on the hadron scale.

\section{Diagrams with bremsstrahlung pion emission}

We have two diagrams with bremsstrahlung pion emission, they are
presented in Fig. \ref{pic_9}. Let us denote the amplitudes in Fig.
\ref{pic_9} as follows:
  \bea
  A^{\rho^+\to\pi^+\pi^0}_\nu (\rm{emission}) =p^{\perp p}_{\pi^+\nu}
  \left[ A(\bigtriangleup^+_0)+A(\bigtriangledown^+_0)+
  A(\bigtriangleup^0_+)+A(\bigtriangledown^0_+)
  \right]\ .
  \label{1-1-20}
  \eea
In the triangle diagrams of Fig. \ref{pic_9} quarks are moving
clockwise.

\begin{figure}[h]
  \centerline{\epsfig{file=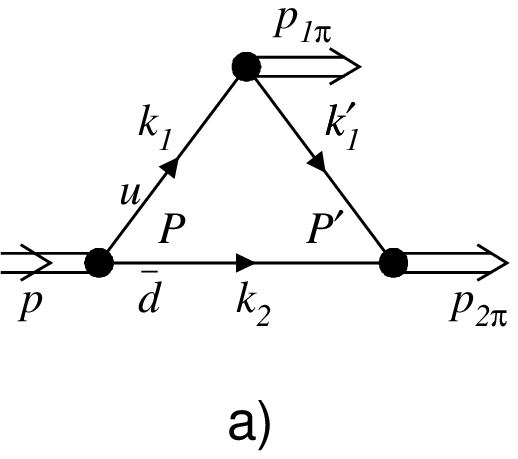,width=4cm}
               \epsfig{file=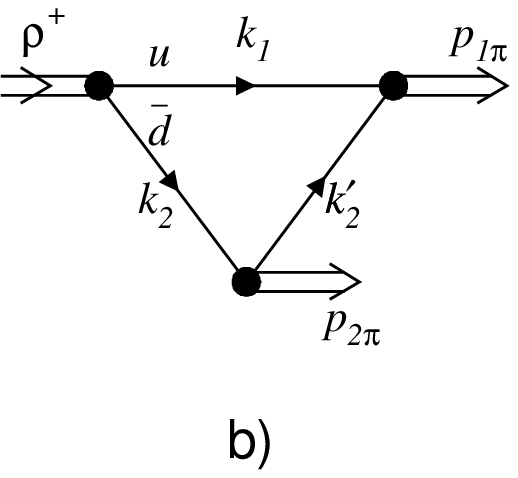,width=4cm}}
  \caption{Processes of pion emission by quarks inside the confinement
  region, with the subsequent $q\bar q\to\pi$ annihilation.
  \label{pic_9}}
  \end{figure}

\subsection{Calculation of the diagrams of Fig. \ref{pic_9}}

Let us first calculate the diagrams in Fig. \ref{pic_9}a,b. In the
framework of spectral integration technics they read:
  \bea
  A(\bigtriangleup^+_0)=\zeta(\bigtriangleup^+_0)\int\limits^\infty_{4m^2}
  \frac{ds}{\pi}
  \frac{ds'}{\pi}\psi_\rho(s)d\Phi_{\triangle}
  (P,P';k_1,k'_1,k_2)g_\pi
  S_{\bigtriangleup^+_0}(s,s',M^2_\pi)\psi_\pi(s'),\nn\\
  A(\bigtriangledown^+_0)=\zeta(\bigtriangledown^+_0)
  \int\limits^\infty_{4m^2}\frac{ds}{\pi}
  \frac{ds'}{\pi}\psi_\rho(s)
  d\Phi_{\nabla}(P,P';k_2,k'_2,k_1)
  g_\pi
  S_{\bigtriangledown^+_0}(s,s',M^2_\pi)\psi_\pi(s').
  \label{1-1-21}
  \eea
Here $\zeta(\bigtriangleup^+_0)$ and $\zeta(\bigtriangleup^+_0)$ are
the isotopic factors,
$\zeta(\bigtriangleup^+_0)=\zeta(\bigtriangleup^+_0)=-1/\sqrt{2}$
(we mean $\pi^0=(u\bar u -d\bar d)/\sqrt{2}$; the phase volume of
the triangle diagram can be written in the standard way in the
double spectral integral:
  \bea
  d\Phi_{\triangle}(P,P';k_1,k'_1,k_2)\frac{1}{64\pi}
  \frac{d^3k_1}{k_{10}} \frac{d^3k_2}{k_{20}} \frac{d^3k'_1}{k'_{10}}
  \delta^{(4)}(P-k_1-k_2) \delta^{(4)}(P'-k'_1-k_2),\nn \\
  d\Phi_{\nabla}(P,P';k_2,k'_2,k_1)\frac{1}{64\pi}
  \frac{d^3k_2}{k_{20}} \frac{d^3k_1}{k_{10}} \frac{d^3k'_2}{k'_{20}}
  \delta^{(4)}(P-k_1-k_2) \delta^{(4)}(P'-k'_2-k_1).
  \label{1-1-22}
  \eea
The spin factors $S_{\bigtriangleup^+_0}(s,s',\mu^2_\pi)$ and
$S_{\bigtriangledown^+_0}(s,s',M^2_\pi)$ are calculated by
considering the traces corresponding to the quark loops of Fig.
\ref{pic_9}a:
  \bea
  &&-Sp[(\hat k_1+m)i\gamma_5(\hat k'_1+m)i\gamma_5 (-\hat
  k_2+m)\gamma^{\perp P}_\nu]= (-P'^{\perp
  P}_\nu)S_{\bigtriangleup^+_0}(s,s',M^2_\pi),
  \nn\\
  &&-Sp[(\hat k_1+m)i\gamma_5(-\hat k'_2+m)i\gamma_5 (-\hat
  k_2+m)\gamma^{\perp P}_\nu]= (-P'^{\perp
  P}_\nu)S_{\bigtriangledown^+_0}(s,s',M^2_\pi).
  \label{1-1-23}
  \eea
We have
  \bea
  S_{\bigtriangleup^+_0}(s,s',M^2_\pi)=
  S_{\bigtriangledown^+_0}(s,s',M^2_\pi) =\frac{8M^2_\pi
  ss'}{-(s-s')^2+2M^2_\pi(s+s')-M^4_\pi}= \frac{-8M^2_\pi
  ss'}{\lambda(s,s',M^2_\pi)}.
  \label{1-1-24}
  \eea
Calculating the spin factor, we gave the definition (\ref{1-1-20})
and used the relation  $p^{\perp p}_{\pi^0\,\nu}=-p^{\perp
p}_{\pi^+\,\nu}$.

In (\ref{1-1-21}) the pion emission constant $g_\pi$ is  present. It
was determined from the reactions $\rho\to\gamma\pi$ and
$\omega\to\gamma\pi$ \cite{g_pi}. We have two solutions:
  \bea
  \label{RD-1}
  {\rm Solution \, I}  &:&\qquad g_\pi=  16.7 \pm 0.3\ ^{+0.1}_{-2.3}\ ,\nn
  \\
  {\rm Solution \, II} &:&\qquad g_\pi=       -3.0  \pm 0.3\ ^{+2.1}_{-0.1}\
  .
  \eea
In Eq. (\ref{RD-1}) we have included systematical errors
($(+0.1/-2.3)$ for Solution I and $(+2.1/-0.1)$ for Solution II)
which are caused by the uncertainties of the fit of $q\bar q$ wave
functions in the spectral integral equation.

So, we have regions of positive and negative $g_\pi$. However, one
should take into account that the sign of $g_\pi$ in (\ref{RD-1}) is
rather conventional: it depends on the signs of the wave functions
of photons and mesons involved in the calculation. Because of that,
to be precise, we should state that for $g_\pi$ we determine
absolute values only.

Solution I gives us the value of the of pion--nucleon coupling;
recall that it is determined as a factor in the phenomenological
Lagrangian: $g_{\pi NN}\bigg(\bar \psi\, '_N
(\vec\tau\vec\varphi_\pi)i\gamma_5\psi_N\bigg)$. It is in agreement
with the results for pion--nucleon scattering $ g_{\pi
 NN}^2/4\pi\simeq 14 $ \cite{stoks,arndt,bugg-g}.

The amplitude $A(\bigtriangleup^+_0)$ can be calculated directly by
(\ref{1-1-24}) if $M^2_\pi < 0$, {\it i.e.} in the non-physical pion
mass region. This is, however, not a serious problem: one can carry
out a few calculations at $M^2_\pi < 0$ close to $M^2_\pi= 0$, and
after that extrapolate, by making use, {\it e.g.}, of the formula
  \be
  A(\bigtriangleup^+_0, \;M^2_\pi)\simeq a+bM^2_\pi+cM^4_\pi
  \ee
to arrive at $M^2_\pi=0.02$ GeV$^2$ (or, $M_\pi=140$ MeV).

Hence, let us write (\ref{1-1-21}) in the form:
  \bea
  &&A(\bigtriangleup^+_0)\!\!=\!\!\zeta(\bigtriangleup^+_0)
  \int\limits^\infty_{4m^2}\frac{dsds'}{\pi^2}
  \psi_\rho(s)\frac{\Theta(-M^2_\pi
  ss'-M^2_\pi\lambda(s,s',M^2_\pi))}
  {16\sqrt{\lambda(s,s',M^2_\pi)}} g_\pi
  S_{\triangle^+_0}(s,s',M^2_\pi)\psi_\pi(s'), \nn \\
  &&A(\bigtriangleup^+_0,M^2_\pi=0)\!\!=\!\!\zeta(\bigtriangleup^+_0)
  \int\limits^\infty_{4m^2}\frac{ds}{16\pi^2}
  \psi_\rho(s)
    g_\pi \psi_\pi(s) 4s\sqrt{1-\frac{4m^2}s}.
  \label{1-1-25-0}
  \eea

Similar calculations demonstrate that the contributions of the
processes of Fig. \ref{pic_9} are equal:
  \bea
  A(\bigtriangleup^+_0)=A(\bigtriangledown^+_0)=
  A(\bigtriangleup^0_+)=A(\bigtriangledown^0_+),
  \label{1-1-26}
  \eea

\subsection{Results of the calculations of the bremsstrahlung emission
  width}

The width of the $\rho\to\pi\pi$ decay is determined as follows:
  \bea
  M_\rho\Gamma^{\rm (emission)}_{\rho\to\pi\pi}&=&\int
  d\Phi_2(p ;p_{\pi^0},p_{\pi^+}) \frac 13 \sum_\nu\
  \bigg | p^{\perp p}_{\pi^{+\nu}} \cdot 4A
  \left(\bigtriangleup^{\pi^+}_{\pi^0}\right)
    \bigg |^2 \nn \\
  &=&\frac 13 \frac 1{16\pi}\sqrt{1-\frac{4M^2_\pi}{M^2_\rho}}
  \left(\frac{M^2_\rho}4 - M^2_\pi\right)\, 16
  \bigg |A \left(\bigtriangleup^{\pi^+}_{\pi^0}\right)\bigg |^2
  \eea
The factor $1/3$ is the consequence of averaging over the $\rho$
meson spin. Recall that $p^2=M^2_\rho$ and $| p^{\perp
p}_{\pi^{+\,\nu}}|^2 = M^2_\rho /4 - M^2_\pi$.

The processes containing only pion emission give widths essentially
larger than those observed in experiments. For Solution I, Eq.
(\ref{RD-1}), we obtain: \\
$\Gamma_{\rho(775)\to\pi\pi}(bremsstrahlung\, emission\, of\, pion)
\simeq 2000$ MeV. \\ This means that the pion bremsstrahlung amplitudes
are reconciled with the amplitude of the direct transition $q\bar
q\to\pi\pi$, Fig. \ref{pic_6}c.

\section{The confinement singularity and the direct transition process
$q\bar q\to\pi\pi$, Fig. \ref{pic_6}c}

Considering the confinement singularity which is the attribute of
the direct $q\bar q\to\pi\pi$ transition, Fig. \ref{pic_6}c, we make
use of the results obtained in \cite{SI-qq} when searching for
$q\bar q$-levels in the framework of the spectral integral approach.

\begin{figure}[h]
\centerline{\epsfig{file=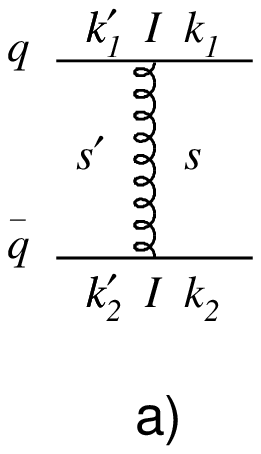,width=4cm}
            \epsfig{file=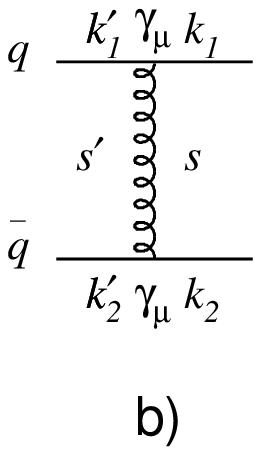,width=4cm}
            \epsfig{file=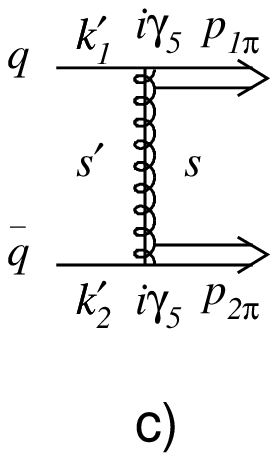,width=4cm}}
   \caption{Amplitude with confinement singularities: a) scalar exchange,
b) vector exchange, c)~quark (fermionic) exchange.
\label{pic_7}}
\end{figure}

\subsection{Confinement singularities}

We discuss here three types of amplitudes with $t$-channel
confinement singularities, they are presented in Fig. \ref{pic_7}.
In Figs. \ref{pic_7}a and \ref{pic_7}b we show $t$-channel
singularities for scalar and vector exchanges between quark and
antiquark; they were used in \cite{SI-qq} when obtaining the $q\bar
q$-levels (see also Fig. \ref{pic_5}d where the corresponding
equation is shown). In Fig. \ref{pic_7}c the amplitude with a
confinement singularity is drawn for the $q\bar q\to\pi_1\pi_2$
transition.

Introducing the momenta of quarks and antiquarks we stress that the
total energies are not conserved in the spectral integrals (just as in
dispersion relations). Hence, in the general case $s\neq s'$.

\subsubsection{Confinement singularities in $q\bar q$ interaction}

We write the amplitudes of Fig. \ref{pic_7}a,b in the following form
\cite{SI-qq}:
\bea \label{1-1-6}
&&{\rm S-exchange}:\quad
\left(\bar\Psi(k'_1)\,I\,\Psi(k_1)\right)G_S(t_\perp)\, I^{(\mu\to
0)}_1(t_\perp) \left(\bar\Psi(-k_2)\,I\,\Psi(k'_2)\right),
\\ \nn
&&{\rm V-exchange}:\quad -\left(\bar\Psi(k'_1)\,\gamma_\mu\,
\Psi(k_1)\right)G_V(t_\perp)\, I^{(\mu\to 0)}_1(t_\perp)
\left(\bar\Psi(-k_2)\,\gamma_\mu\,\Psi(k'_2)\right).
\eea
 The singular block, $I^{(\mu\to 0)}_1(t_\perp)$, is given in
(\ref{1-1-5}), where $t_\perp=(q^\perp)^2=(k_1^{\perp p}-k'^{\perp
p'}_1)^2$ with $p=k_1+k_2$ and $p'=k'_1+k'_2$ ($p^2=s$, $p'^2=s'$);
the transverse components are defined as $k^{\perp
p}_{1\alpha}=g^{\perp p}_{\alpha\beta}k_{1\beta}$, $g^{\perp
p}_{\alpha\beta}=g_{\alpha\beta}-p_\alpha p_\beta/p^2$ and
$k'^{\perp p'}_{1\alpha}=g^{\perp p'}_{\alpha\beta}k'_{1\beta}$,
$g^{\perp p'}_{\alpha\beta}=g_{\alpha\beta}-p'_\alpha p'_\beta
/p'^2$.

In \cite{SI-qq} we have used several singular terms corresponding to
the expansion:
\be \label{1-1-6exp}
G_S(t_\perp)=G_S(0)+G_S(1)\,
t_\perp +G_S(2)t^2_\perp, \quad G_V(t_\perp)=G_V(0)+G_V(1)\, t_\perp
+G_V(2)t^2_\perp .
\ee
We use (\ref{1-1-6}) as a pattern in writing
the amplitude for the $q\bar q\to \pi\pi$ transition.

\subsubsection{Confinement singularities in $q\bar q\to \pi\pi$ transition}

The quark-exchange singularity of Fig. \ref{pic_7}c reads:
\be
\label{1-1-6G}
 \phi^*_{\pi(1)}\phi^*_{\pi(2)}\left(\bar\Psi(-k'_2)\, i\gamma_5
 \sum_a\Psi_{conf}^{(a)}(q^\perp)
G_{q\bar q\to \pi\pi}(t_\perp)\,   I^{(\mu\to 0)}_N(t_\perp)\,
\bar\Psi_{conf}^{(a)}(q^\perp) \, i\gamma_5\Psi(k'_1)\right)
\ee
 In Fig. \ref{pic_7}c we have $q^{\perp}=k'^{\perp p'}_1-p_{\pi
1}^{\perp p}$ with $p_{\pi 1\alpha}^{\perp p}=g^{\perp
p}_{\alpha\beta}p_{\pi 1\beta}$.

Below we consider two types of the spin structure for the
confinement singularity.

{\bf (i) Gribov quark-exchange interaction }

Following \cite{Gribov}, we accept that the mass of the $t$-channel
quark state is very small. We put it to be zero. Then we write the
completeness condition in a standard form:
\be \label{1-1-6c}
\sum_a\Psi_{conf}^{(a)}(q^\perp)
    \bar\Psi_{conf}^{(a)}(q^\perp)=\hat q^\perp\, .
\ee
Equation (\ref{1-1-6G}) reads:
\be \label{1-1-6GG2}
\phi^*_{\pi(1)}\phi^*_{\pi(2)}\left(\bar\Psi(-k'_2)\,
   i\gamma_5
\hat q^\perp  I^{(\mu\to 0)}_N(t_\perp)G_{q\bar q\to \pi\pi}(t_\perp)\,
 i\gamma_5
\Psi(k'_1)\right).
\ee
Considering the performed here calculations
as first estimations, we use interaction singularities $I^{(\mu\to
0)}_1(t_\perp)$ with $N=1,0, -1$ and constant non-singular
coefficients $G_{q\bar q\to \pi\pi}(t_\perp)=const$:
\bea
\label{1-1-6cc}
&&G_{q\bar q\to \pi\pi}(t_\perp)\, I^{(\mu\to
0)}_1(t_\perp) \to \sum\limits_{N=-1}^1
G_{q\bar q\to \pi\pi}(1-N)\, I^{(\mu\to 0)}_{N}(t_\perp)\\
&&\qquad= G_{q\bar q\to \pi\pi}(0)\, I^{(\mu\to 0)}_1(t_\perp)+ G_{q\bar
q\to \pi\pi}(1)\, I^{(\mu\to 0)}_0(t_\perp) + G_{q\bar q\to
\pi\pi}(2)\, I^{(\mu\to 0)}_{-1}(t_\perp).\nn
\eea
 Thus, we use the singularities of the same type as in the $q\bar q $
interaction, see (\ref{1-1-6}) and (\ref{1-1-6exp}).


{\bf (ii) Universal confinement singularity }

The universality of the confinement singularity
appears when the spin functions in the $t$-channel exchange obey the
following completeness condition:
\be \label{1-1-6c2}
\sum_a\Psi_{conf}^{(a)}(q^\perp)
    \bar\Psi_{conf}^{(a)}(q^\perp)= I,
\ee
where $ I=diag (1,1,1,1)$ is a unit four-dimensional matrix in
the spin space. We re-write (\ref{1-1-6G}):
\bea \label{1-1-6GG}
&\,&\phi^*_{\pi(1)}\phi^*_{\pi(2)}\left(\bar\Psi(-k'_2)\,
  i\gamma_5  I^{(\mu\to 0)}_1(t_\perp)
\, i\gamma_5 G_{q\bar q\to \pi\pi}(t_\perp) \Psi(k'_1)\right) \\
&=&
-\phi^*_{\pi(1)}\phi^*_{\pi(2)}\left(\bar\Psi(-k'_2)
    I^{(\mu\to 0)}_1(t_\perp)
 G_{q\bar q\to \pi\pi}(t_\perp) \Psi(k'_1)\right),\nn \\
&\,& I^{(\mu\to 0)}_1(t_\perp) G_{q\bar q\to \pi\pi}(t_\perp)\to
\sum\limits_{N=-1}^1 G_{q\bar q\to \pi\pi}(1-N)\, I^{(\mu\to
0)}_{N}(t_\perp).\nn
\eea
 Then Eq. (\ref{1-1-6GG}) gives the singularity of the same type
as in transition $q\bar q \to q\bar q$, see Eq. (\ref{1-1-6}).

\subsubsection{An example: expressions for $q_\perp$ and $t_\perp$ in the
centre-of-mass system}

To make our calculations more understandable, let us present
here the expressions for $q_\perp$ and $t_\perp$ in the
centre-of-mass system, taking into account that in a spectral
integral the initial and final state energies are different.

The momenta in the diagrams Fig. \ref{pic_7}a,b have the following
form in the centre-of-mass frame:
\bea
&&k_1=\left(\frac{\sqrt{s}}{2},\vec
n\sqrt{\frac{s}{4}-m^2}\right),\qquad
k'_1=\left(\frac{\sqrt{s'}}{2},\vec
n'\sqrt{\frac{s'}{4}-m^2}\right),
\\ \nn
&&k_2=\left(\frac{\sqrt{s}}{2},-\vec
n\sqrt{\frac{s}{4}-m^2}\right),\qquad
k'_2=\left(\frac{\sqrt{s'}}{2},-\vec
n'\sqrt{\frac{s'}{4}-m^2}\right).
\label{1-1-8}
\eea
   In accordance with this,
\bea
&&k^\perp_1=\left(0,\vec n\sqrt{\frac{s}{4}-m^2}\right),\qquad
k'^{\perp}_1=\left(0,\vec n'\sqrt{\frac{s'}{4}-m^2}\right),
\\ \nn
&&k^\perp_2=\left(0,-\vec n\sqrt{\frac{s}{4}-m^2}\right),\qquad
k'^{\perp}_2=\left(0,-\vec n'\sqrt{\frac{s'}{4}-m^2}\right),
\label{1-1-9}
\eea
   and in the c.m. system we write
\bea
q^\perp=k^\perp_1-k'^{\perp}_1=-k^\perp_2+k'^\perp_2 =\left(0,\vec
n\sqrt{\frac{s}{4}-m^2}-\vec n'\sqrt{\frac{s'}{4}-m^2}\right).
\label{1-1-10}
\eea
   Similarly, for the process in Fig. \ref{pic_7}c we have:
\bea
&&p_{1\pi}=\left(\frac{s+M^2_{1\pi}-M^2_{2\pi}}{2\sqrt{s}}, \,\,\,\,\vec
n_\pi\sqrt{\frac1{4s}\left[s-(M_{1\pi}+M_{2\pi})^2\right] \nn
\left[s-(M_{1\pi}-M_{2\pi})^2\right]}\right),
\\ \nn
&&p_{2\pi}=\left(\frac{s-M^2_{1\pi}+M^2_{2\pi}}{2\sqrt{s}}, -\vec
n_\pi\sqrt{\frac1{4s}\left[s-(M_{1\pi}+M_{2\pi})^2\right]
\left[s-(M_{1\pi}-M_{2\pi})^2\right]}\right),
\\ \nn
&&p^\perp_{1\pi}=\left(0, \,\,\,\,\vec
n_\pi\sqrt{\frac1{4s}\left[s-(M_{1\pi}+M_{2\pi})^2\right]
\left[s-(M_{1\pi}-M_{2\pi})^2\right]}\right),
\\
&&p^\perp_{2\pi}=\left(0, -\vec
n_\pi\sqrt{\frac1{4s}\left[s-(M_{1\pi}+M_{2\pi})^2\right]
\left[s-(M_{1\pi}-M_{2\pi})^2\right]}\right).
\label{1-1-11}
\eea
that leads to
\bea
q^\perp= \left(0,\vec n\sqrt{\frac{s}{4}-m^2} -\vec
n_\pi\sqrt{\frac1{4s}\left[s-(M_{1\pi}+M_{2\pi})^2\right]
\left[s-(M_{1\pi}-M_{2\pi})^2\right]} \right) .
\label{1-1-12}
\eea
 In order to preserve the possibility to generalize,
we take into account in (\ref{1-1-11}) and (\ref{1-1-12}) that
$\pi_1$ and $\pi_2$ can have different masses, respectively,
$M_{1\pi}$ and $M_{2\pi}$. Below we neglect this difference and put
$M_{1\pi}=M_{2\pi}\equiv M_{\pi}$.

\begin{figure}[h]
\centerline{\epsfig{file=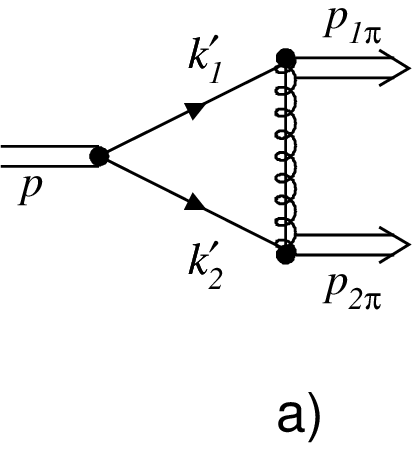,width=4cm}
            \epsfig{file=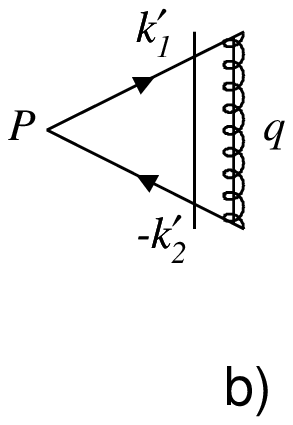,width=4cm}}
\caption{Triangle diagram $\rho\to\pi^+\pi^0$, and the notation of
momenta in the spectral integral.
\label{pic_8}}
\end{figure}

\subsection{Calculation of the direct transition amplitude  $\rho\to\pi\pi$
in the framework of spectral integration technics}

Here we calculate amplitude of the transition Fig. \ref{pic_6}c; the
transition diagram with the notation of momenta is shown separately
in Fig. \ref{pic_8}a.

The amplitude of the $\rho^+\to\pi^+\pi^0$ decay has the structure
\bea
  A_\nu\left(\triangleleft^{\pi^+}_{\pi^0};s,M^2_\rho\right)+
  A_\nu\left(\triangleleft^{\pi^0}_{\pi^+};s,M^2_\rho\right)&=&
p^\perp_{\pi^+\nu}\, \bigg [
A\left(\triangleleft^{\pi^+}_{\pi^0};s,M^2_\rho\right)+
  A\left(\triangleleft^{\pi^0}_{\pi^+};s,M^2_\rho\right)\bigg ]\nn\\
&=&p^\perp_{\pi^+\nu}\,
A(\triangleleft;s,M^2_\rho)({\rm direct\, transition})\ .
\label{1-1-13n}
\eea
  Let us remind that
$p^{\perp p}_{\pi^+\nu}=-p^{\perp p}_{\pi^0\nu}$ and
$p=p_{\pi^+}+p_{\pi^0}$. In spectral integrals the energy is not
conserved, so $s\neq M^2_\rho$. For physical amplitudes $s=
M^2_\rho$.

To be definite, let us consider
$A_\nu\left(\triangleleft^{\pi^+}_{\pi^0};s,M^2_\rho\right)$.

\subsubsection{ Gribov quark-exchange interaction }

For the Gribov quark-exchange interaction the amplitude reads:
\bea
\label{1-1-15n}
&&A_\nu\left(\triangleleft^{\pi^+}_{\pi^0};s,M^2_\rho\right)=
\int\limits^\infty_{4m^2} \frac{ds'}{\pi} \Psi_\rho(s')
d\Phi_2(p';k'_1,k'_2)\\
&&\times (-) Sp\bigg[(\hat k'_1 +m)\hat { q}^\perp
 G_{q\bar q\to \pi\pi}(t_\perp)    I^{(\mu\to 0)}_N(t_\perp)
(-k'_2 +m) \gamma_\nu^{\perp p'}\bigg]. \nn
\eea
 Here the $\rho$-meson wave function is the ratio of the vertex of
$\rho\to\pi\pi$, $G_{\rho\to q\bar q}(s')$ and the dispersion
integral denominator $\Psi_\rho(s)=G_{\rho\to q\bar
q}(s')/(s'-M^2_\rho)$, while the two-particle phase volume is
written in a standard form:
\bea
d\Phi_2(p';k'_1,k'_2)=\frac12
(2\pi)^4\frac{d^3k'_1}{2k'_{10}(2\pi)^3}
\frac{d^3k'_2}{2k'_{20}(2\pi)^3}\delta^4(p'-k'_1-k'_2) \,
\label{1-1-17} \eea
  with $p'^2=s'$, $k'^2_1=m^2$, $k'^2_2=m^2$. The transverse momentum
is $ q^\perp= k'^\perp_1 -p^\perp_{\pi^+}$, or, in the c.m.
system (see section 2.1.3):
$q^\perp= (0,\vec n'\sqrt{s'/4-m^2} -\vec n_\pi\sqrt{s/4-M^2_\pi} )$.

To extract the invariant part of the amplitude $p^\perp_{\pi^+\nu}\,
A\left(\triangleleft^{\pi^+}_{\pi^0};s,M^2_\rho\right)$, we write:
\bea \label{1-1-15nn}
&&A\left(\triangleleft^{\pi^+}_{\pi^0};s,M^2_\rho\right)=
\int\limits^\infty_{4m^2}
\frac{ds'}{\pi} \Psi_\rho(s')
d\Phi_2(p';k'_1,k'_2) \\
&& \times(-) Sp\bigg[(\hat k'_1 +m)\hat { q}^\perp
 G_{q\bar q\to \pi\pi}(t_\perp)   I^{(\mu\to 0)}_1( t_\perp)
(-\hat k'_2 +m)
\gamma_\nu^{\perp p'}\bigg]
\frac{p^{\perp p}_{\pi^+\, \nu}}{(p^{\perp p}_{\pi^+\, \nu'})^2}      \nn
\\
&& =\int\limits^\infty_{4m^2}
\frac{ds'}{\pi}
\Psi_\rho(s')
d\Phi_2(p';k'_1,k'_2)
(-8)\frac{4(k'_1q^\perp)(k'_1p_{1\pi})+s'(q^\perp p_{1\pi})}
{4M_\pi^2- s} \, ,\nn
\eea
where
$(k'_1p_{1\pi})=-z'\sqrt{s'/4-m^2}\sqrt{s/4-M_\pi^2}\,$,
$\quad (k'_1q^\perp)=-s'/4+m^2+z'\sqrt{s'/4-m^2}\sqrt{s/4-M_\pi^2}\,$ and
$(q^\perp p_{1\pi})=s/4-M_\pi^2-z'\sqrt{s'/4-m^2}\sqrt{s/4-M_\pi^2}\,$.

It is easy to see that in (\ref{1-1-13n}) we have
$A\left(\triangleleft^{\pi^+}_{\pi^0};s,M^2_\rho\right)=
A\left(\triangleleft^{\pi^0}_{\pi^+};s,M^2_\rho\right)$ and, hence,
\be
  p^\perp_{\pi^+\nu}\, A(\triangleleft;s,M^2_\rho)=  p^\perp_{\pi^+\nu}
  \cdot 2 A\left(\triangleleft^{\pi^+}_{\pi^0};s,M^2_\rho\right).
  \label{1-1-13z}
\ee
  Let us emphasize once more that the amplitude
$A_\nu\left(\triangleleft^{\pi^+}_{\pi^0};s,M^2_\rho\right)$ is an
auxiliary one, the physical amplitude requires $s\to M^2_\rho$:
\be
\bigg
[A_\nu\left(\triangleleft^{\pi^+}_{\pi^0};s,M^2_\rho\right)\bigg
]_{s\to M^2_\rho} =A_\nu\left(\triangleleft^{\pi^+}_{\pi^0};M^2_\rho
,M^2_\rho\right)\ .
\ee

\subsubsection{ Universal confinement singularity }

To write the amplitude with universal confinement interaction, we
should change in (\ref{1-1-15nn}) the $t$-channel confinement block:
$(\hat { q}^\perp ) I^{(\mu\to 0)}_1( t_\perp) \,
 G_{q\bar q\to \pi\pi}(t_\perp)
\to  (-)   I^{(\mu\to 0)}_1( t_\perp) G_{q\bar q\to \pi\pi}(t_\perp)$.
Therefore
\bea
\label{1-1-15uni}
&&A\left(\triangleleft^{\pi^+}_{\pi^0};s,M^2_\rho\right)
=\int\limits^\infty_{4m^2}
\frac{ds'}{\pi}\Psi_\rho(s')d\Phi_2(p';k'_1,k'_2)
\\
&&  \times Sp\bigg[(\hat k'_1 +m)
 G_{q\bar q\to \pi\pi}(t_\perp)   I^{(\mu\to 0)}_1( t_\perp)
(-k'_2 +m)
\gamma_\nu^{\perp p'}\bigg]
\frac{p^{\perp p}_{\pi^+\, \nu}}{(p^{\perp p}_{\pi^+\, \nu'})^2}
\nn \\
&&=\int\limits^\infty_{4m^2}
\frac{ds'}{\pi} \Psi_\rho(s') d\Phi_2(p';k'_1,k'_2)
G_{q\bar q\to\pi\pi}(t_\perp)   I^{(\mu\to 0)}_1( t_\perp)
\left[8mz'\frac{\sqrt{s'/4-m^2}}{\sqrt{s/4-M_\pi^2}}\right]\, .
\nn
\eea
In the right-hand side of (\ref{1-1-15uni}) we put, as
previously, $ G_{q\bar q\to \pi\pi}(t_\perp)=G_{q\bar q\to
\pi\pi}(0)+G_{q\bar q\to \pi\pi}(1)\,
  t_\perp +G_{q\bar q\to \pi\pi}(2)t^2_\perp $.

\subsection{Estimation of confinement
interaction couplings $G_{q\bar q\to \pi\pi}(n)$ }

Taking into account both the bremsstrahlung type and the direct
confinement interaction transitions, the width of the
$\rho\to\pi\pi$ decay is determined as follows:
\bea \label{G1}
M_\rho\Gamma_{\rho\to\pi\pi}&=&\int d\Phi_2(p ;p_{\pi^0},p_{\pi^+})
\frac 13 \sum_\nu\ \bigg | A^{\rho^+\to\pi^+\pi^0}_\nu
(\rm{emission})+ A^{\rho^+\to\pi^+\pi^0}_\nu (\rm{direct\,
transition})
 \bigg |^2 \nn \\
&=&\frac 13 \frac 1{16\pi}\sqrt{1-\frac{4M^2_\pi}{M^2_\rho}}
\left(\frac{M^2_\rho}4 - M^2_\pi\right)\,
\bigg |4A \left(\bigtriangleup^{\pi^+}_{\pi^0};M^2_\rho,M^2_\rho\right)+2
A\left(\triangleleft^{\pi^+}_{\pi^0};M^2_\rho,M^2_\rho\right)\bigg |^2\ .
\eea
 The factor $1/3$ is the consequence of averaging over the $\rho$
meson spin. Let us recall that here
 $p^2=M^2_\rho$ and $| p^{\perp p}_{\pi^{+}\nu}|^2=
 M^2_\rho /4 - M^2_\pi$.

\subsubsection{Order of values of $G_{q\bar q\to \pi\pi}(1-N)$
at $N=0,1$}

To estimate the order of values of $G_{q\bar q\to \pi\pi}(1-N)$ in
(\ref{1-1-6cc}) and (\ref{1-1-6GG}), we fit the $\rho$-meson width
($\Gamma_{\rho\to\pi\pi}=150$ MeV) which is defined according Eq.
(\ref{G1}), using one non-zero coupling $G_{q\bar q\to \pi\pi}(1-N)$
only. Each value of $g_\pi$ results in two values of $G_{q\bar q\to
\pi\pi}(1-N)$ because of the constructive and destructive
interferences of $4A
\left(\bigtriangleup^{\pi^+}_{\pi^0};M^2_\rho,M^2_\rho\right)$ and
$2A\left(\triangleleft^{\pi^+}_{\pi^0};M^2_\rho,M^2_\rho\right)$ in
(\ref{G1}).
 We obtain for the Gribov quark-exchange and for the universal interactions
the following couplings (in GeV units).\\
{\bf (i) Gribov quark-exchange interaction}:
\begin{equation} \label{G2}
{\begin{tabular}{l|c|c}
   $g_\pi$ & $G_{q\bar q\to \pi\pi}(0)$, $N=1$ & $G_{q\bar q\to \pi\pi}(1)$, $N=0$
                          \\
\hline
$-2.96$     & $(0.016\,\,{\rm or}\,\,0.112)$    &$0 $
                      \\
$16.85$     & $ (-0.411\,\,{\rm or}\,\,-0.315)$ &$0  $
                        \\
\hline
$-2.96$     & $ 0   $&$ (0.076\,\,{\rm or}\,\,0.532) $
                      \\
$16.85$     & $ 0 $  &$ (-1.957\,\,{\rm or}\,\,-1.501)$
                        \\
\end{tabular} }
\end{equation}

{\bf (ii)} Universal confinement singularity:
\begin{equation} \label{G3}
{\begin{tabular}{l|c|c}
   $g_\pi$ & $G_{q\bar q\to \pi\pi}(0)$, $N=1$ & $G_{q\bar q\to \pi\pi}(1)$, $N=0$
                          \\
\hline
$-2.96$     & $(0.007\,\,{\rm or}\,\,0.047)$    &$0 $
                      \\
$16.85$     & $ (-0.172\,\,{\rm or}\,\,-0.132)$ &$0  $
                        \\
\hline
$-2.96$     & $ 0   $&$ (0.112\,\,{\rm or}\,\,0.784) $
                      \\
$16.85$     & $ 0 $  &$ (-2.886\,\,{\rm or}\,\,-2.213)$
                        \\
\end{tabular} }
\end{equation}

\section{Self-energy part in the $\rho\to\pi\pi\to\rho$ transition}

\begin{figure}[h]

\centerline{\epsfig{file=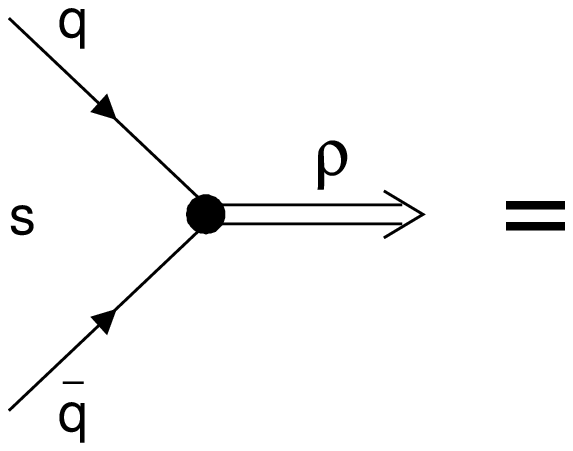,height=3.5cm}
            \epsfig{file=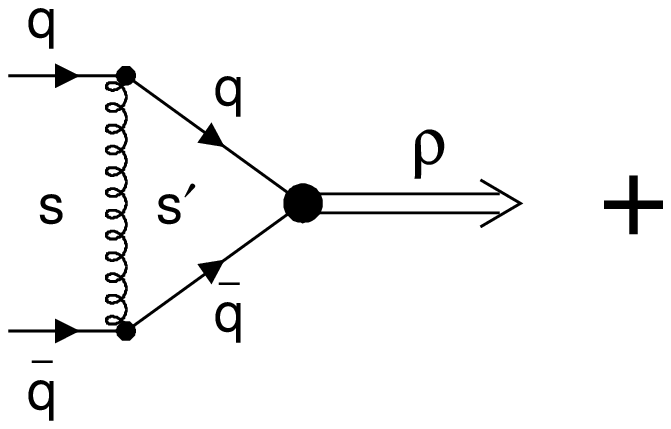,height=3.5cm}
            \epsfig{file=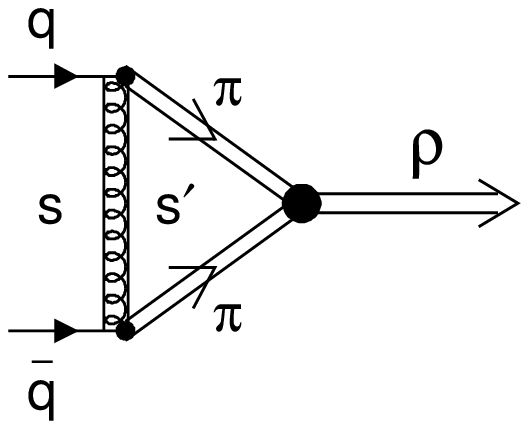,height=3.5cm}}
\vspace{-5mm}
\centerline{\epsfig{file=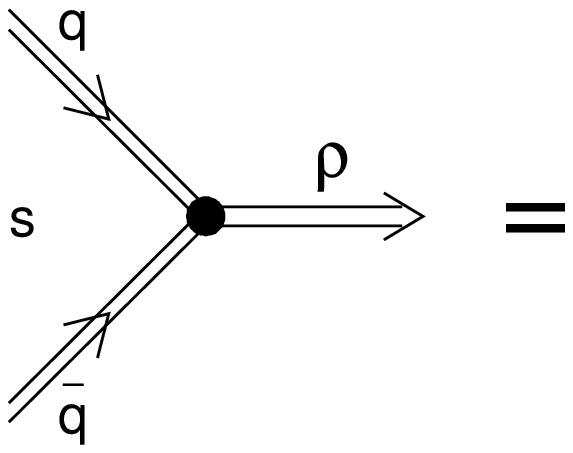,height=3.5cm}
            \epsfig{file=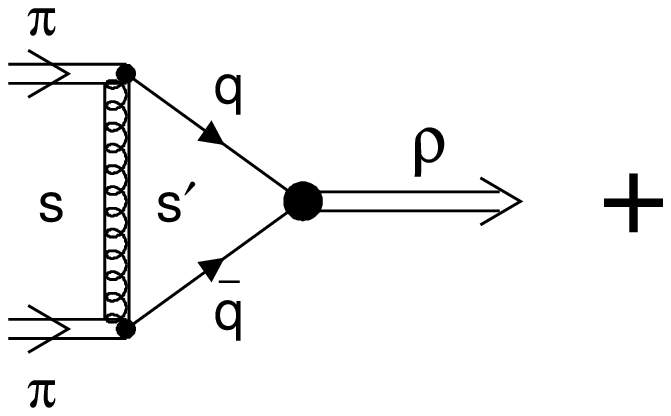,height=3.5cm}
            \epsfig{file=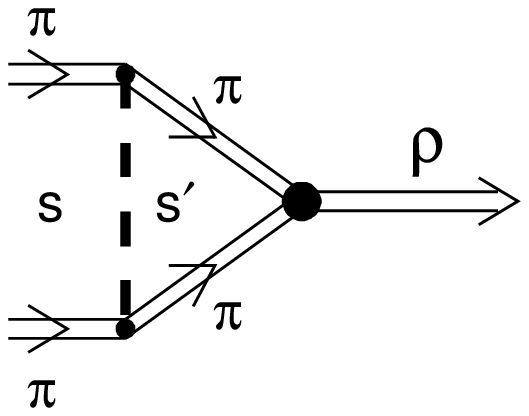,height=3.5cm}}
\vspace{-5mm} \caption{Graphical representation of two-channel
($q\bar q$, $\pi\pi$) equations for the $\rho$-meson.
\label{qq-pipi-eq}}
\end{figure}

\begin{figure}[h]
\centerline{\epsfig{file=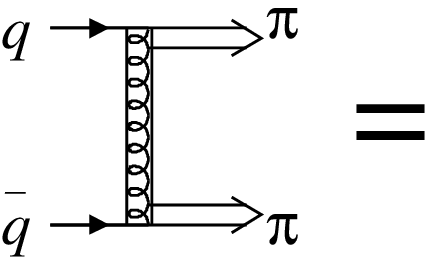,height=3.5cm}
            \epsfig{file=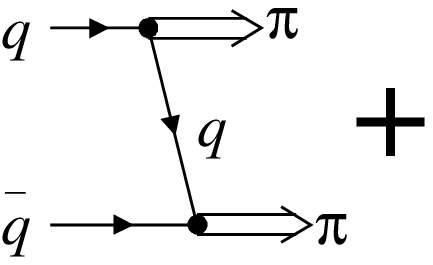,height=3.5cm}
            \epsfig{file=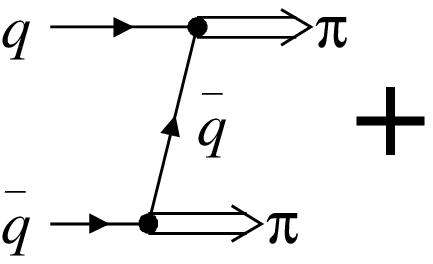,height=3.5cm}
            \epsfig{file=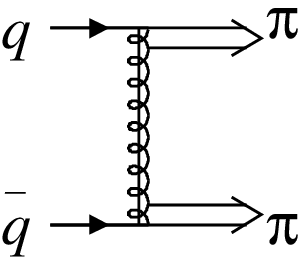,height=3.5cm}}
\vspace{-5mm}
\caption{Three terms of the transition amplitude $q\bar q\to \pi\pi$.
\label{qq-pipi-eq1}}
\end{figure}

In \cite{SI-qq} we have calculated the levels and wave functions of
states in the infinite wall approach (the requirement $\mu\to 0$ in
Eq. (\ref{1-1-2})). Generally speaking, this means that in the width
calculated according to Eq. (\ref{G1}) we neglect the reverse
influence of the $\pi\pi$ channel on the characteristics of the
$\rho$-meson. This is not a bad approximation for the $\rho$-meson,
but in our consideration we can make the next step -- to take into
account the $\pi\pi$ channel -- easily. This two-component equation
is shown in a graphical form in Fig. \ref{qq-pipi-eq}: one component
refers to the $(\rho\to q\bar q)$-vertex, the second one to the
$(\rho\to \pi\pi )$-vertex. The interaction block $ q\bar q\to
\pi\pi$ is deciphered in Fig. \ref{qq-pipi-eq1}: it contains
transitions considered in Chapters 2 and 3.

\begin{figure}[h]
\centerline{\epsfig{file=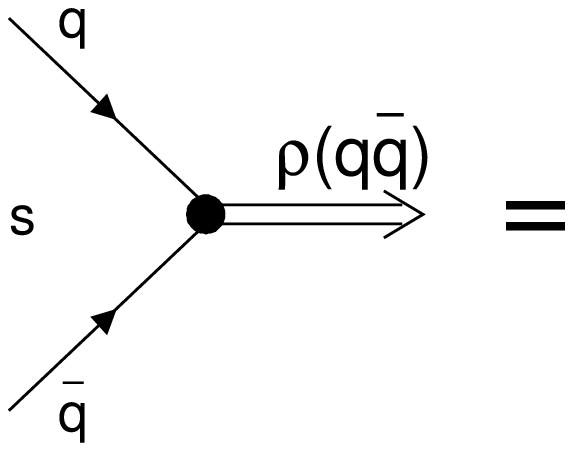,height=3.5cm}
            \epsfig{file=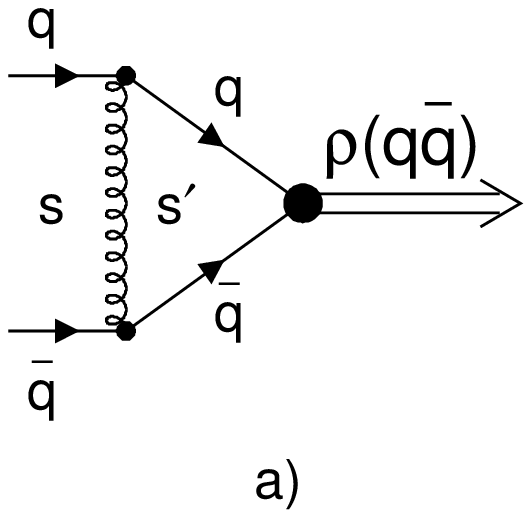,height=3.5cm}}
\vspace{-5mm} \centerline{\epsfig{file=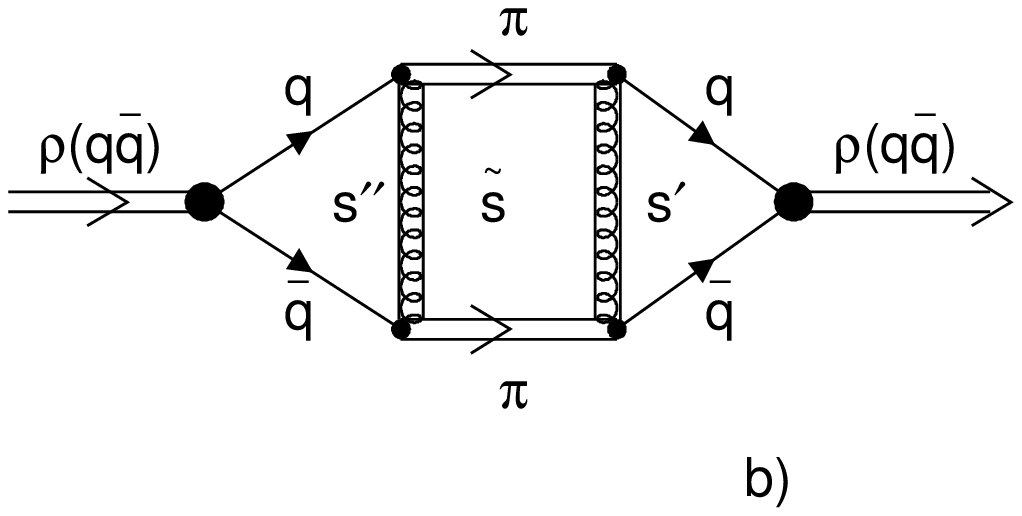,height=3.5cm}}
\caption{Graphical representation of a) the equation for the pure
$q\bar q$ state $\rho(q\bar q)$ and b) the self-energy part
$B(s,M^2_{\rho(q\bar q)})$ which determines the admixture of the
$\pi\pi$ component in the $\rho$-meson according to Eq. (\ref{G4}).
\label{qq-pipi-eq2}
 }
\end{figure}

Two simplifying constraints should be taken into account:\\
(i) the $\pi\pi$ interaction in the $\rho$-meson region is small,\\
(ii) the $\rho$-meson width is not large.\\
Then we have a standard one-channel equation for $\rho_{(q\bar q)}$,
where $\rho_{(q\bar q)}$ is a pure $q\bar q$-state, see Fig.
\ref{qq-pipi-eq2}a, while the $\pi\pi$ channel reveals itself in the
self-energy part of transition $\rho_{(q\bar q)}\to \pi\pi\to
\rho_{(q\bar q)}$ only, see Fig. \ref{qq-pipi-eq2}b. We denote this
self-energy part as $B(s,M^2_{\rho(q\bar q)})$ with the initial and
final state energy squared $M^2_{\rho(q\bar q)}$ and $s$.

The propagator of the pure $q\bar q$-state is transformed as follows:
\bea \label{G4}
\frac{\sum\limits_a \epsilon^{(a)}_\nu
\epsilon^{(a)+}_{\nu'}}{M^2_{\rho(q\bar q)}-s}
&\to &\frac{\sum\limits_a \epsilon^{(a)}_\nu
\epsilon^{(a)+}_{\nu'}}{M^2_{\rho(q\bar q)}-s-B(s,M^2_{\rho(q\bar q)})}\\
&=& \frac{\sum\limits_a \epsilon^{(a)}_\nu \epsilon^{(a)+}_{\nu'}}
{\bigg(M^2_{\rho(q\bar q)}-{\rm Re}B(s,M^2_{\rho(q\bar q)})\bigg)
-s-i\,{\rm Im}B(s,M^2_{\rho(q\bar q)})}\, .\nn
\eea
 The pole requirement reads
 \be \label{G5}
 M^2_{\rho(q\bar q)}-{\rm
Re}B(s,M^2_{\rho(q\bar q)})-s=0\quad {\rm at} \quad
s=M^2_{\rho}\equiv (0.775)^2\, {\rm GeV}^2 .
\ee
 Then the width $\Gamma_{\rho}$ is equal to:
 \be \label{G6}
{\rm Im}B(M^2_{\rho},M^2_{\rho(q\bar q)})=M_{\rho}\Gamma_{\rho}\, .
\ee From the results of the fit \cite{SI-qq} we know that
$M^2_{\rho}= M^2_{\rho(q\bar q)}$, {\it i.e.} the fit provides us
with a negligibly small value of the mass shift. So, we may accept:
\be \label{G8}
{\rm Re}\,B(M^2_{\rho},M^2_{\rho(q\bar q)})= 0.
\ee

\subsubsection{Self-energy part $B(s,M^2_{\rho(q\bar q)})$}

The self-energy part of a pure $q\bar q$-state reads
\be \label{G10}
B(s,M^2_{\rho(q\bar q)})=\int\limits_{4M^2_\pi}^\infty \frac{d\tilde
s}{\pi}\frac{{\rm Im}B(\tilde s,M^2_{\rho(q\bar q)})}{\tilde
s-s-i0},
\ee
where
\be \label{G10Im}
{\rm Im}B( s,M^2_{\rho(q\bar
q)})=\frac{1}{192\pi} \sqrt{\frac{( s-4M^2_\pi)^3}{ s}} \bigg (4A
\left(\bigtriangleup^{\pi^+}_{\pi^0}; s,M^2_\rho\right)+2
A\left(\triangleleft^{\pi^+}_{\pi^0}; s,M^2_\rho\right)\bigg )^2\, .
\ee
Here we take into account that the amplitudes
$A\left(\bigtriangleup^{\pi^+}_{\pi^0}; s,M^2_\rho\right)$ and
$A\left(\triangleleft^{\pi^+}_{\pi^0}; s,M^2_\rho\right)$ do not
 have imaginary parts in the $\mu\to 0$ limit despite of the fact
that $q\bar q$ is present in their intermediate states, see Fig.
\ref{pic_6} and Eqs. (\ref{1-1-21}), (\ref{1-1-15nn}),
(\ref{1-1-15uni}). This means non-flying out quarks in the resonance
decay, {\it i.e.} quark confinement. The only particles flying out
are pions; this fact manifests itself in the presence of the
threshold singularity in (\ref{G10}) at $ s=4M^2_\pi$.

The requirement $M^2_{\rho}= M^2_{\rho(q\bar q)}$ given in Eq.
(\ref{G8}) reads:
\be \label{G11}
{\rm P}\,
\int\limits_{4M^2_\pi}^\infty \frac{d\tilde s}{\pi} \frac{{\rm
Im}B(\tilde s,M^2_{\rho(q\bar q)})}{\tilde s-M^2_{\rho(q\bar q)}}
=0,
\ee
here ${\rm P}$ means the principal value of the integral.

It is natural to think that the requirement (\ref{G11}) is realized
by the behaviour of ${\rm Im}B(\tilde s,M^2_{\rho(q\bar q)})$ at
large $\tilde s$ but not in the region $\tilde s\sim M^2_{\rho(q\bar
q)}$. The region of large $\tilde s$ is, however, beyond our present
considerations.
We could try to produce the condition (\ref{G8}) staying in
the framework of our calculations (by, {\it e.g.}, choosing the
couplings for $G_{q\bar q\to \pi\pi}(0)$, $G_{q\bar q\to
\pi\pi}(1)$, $G_{q\bar q\to \pi\pi}(2)$, or by introducing a
cut-off, or by combining these methods). These would be, however,
artificial procedures, not corresponding to the real formation
mechanism of the self-energy part. Hence, we will not consider this
possibility.

\section{Conclusion}

In the quark model the decay of a $q\bar q$ state into pions ({\it
i.e.} when the quarks fly out of their ``trap'') is described by
processes shown in Fig. \ref{pic_6}. Let us underline that, from the
point of view of a quantum mechanical or a dispersion description,
where time ordering exists, these are different processes -- not so
in the Feynman diagram ideology which does not include time
ordering. We consider the problem in the framework of the dispersion
technics approach.

Our investigations show that the bremsstrahlung-type radiation of
pions ({\it i.e.} radiation coming from the region of the
confinement trap) is rather large. If this were the only possible
process, it would lead to a very broad decay width of the
$\rho$-meson, $\sim 2000$ MeV. But the process which involves the
Gribov singularity, Fig \ref{pic_6}, prevents such a ``smearing''
and we see quite a number of relatively narrow $q\bar q$ states.

Let us note that this is not the only process leading to narrow
$q\bar q$ states. The accumulation of widths of highly excited
$q\bar q$-resonances by their exotic neighbours ({\it e.g.} by
glueballs, see \cite{book3} -- Chapter 2 and references therein)
leads to a similar effect of keeping the quarks inside the trap.
But, and this is important to underline, such an accumulation of
widths can take place only in those regions and for those states
where exotic neighbours exist, while the process with Gribov
singularities is always possible, {\it i.e.} it is universal.

The Gribov singularities are not determined unambiguously from the
investigation of the $\rho(775)\to \pi\pi$ decay. Their
determination is, apparently, possible, but this requires a much
more complete knowledge of the $\pi\pi$ decays.

Quark confinement, {\it i.e.} the absence of $q\bar q$-singularities
in hadron amplitudes when there are $q\bar q$-systems in the
intermediate states is owing to interactions shown in Fig.
\ref{pic_7}a and Fig. \ref{pic_7}b. But the physics of $q\bar
 q$-resonances has to deal also with the interactions in Fig.
\ref{pic_7}c. Indeed, the interactions Fig. \ref{pic_7}a and Fig.
\ref{pic_7}b guarantee the confinement of quarks (at $\mu\to 0$),
but can not prevent a very fast decay, a ``smearing'' of a $q\bar
q$-system.

   \end{document}